\documentclass[]{spie}   
\usepackage{graphicx}

\title{Platform Deformation Refined Pointing and Phase Correction for the AMiBA Hexapod Telescope}

\author{Patrick Koch\supit{a}, Michael Kesteven\supit{b}, Yu-Yen Chang\supit{a}, 
Yau-De Huang\supit{a}, Philippe Raffin\supit{a},
 Ke-Yung Chen\supit{a}, Guillaume Chereau\supit{a}, Ming-Tang Chen\supit{a}, 
Paul.T.P.Ho\supit{a}, Chih-Wie Huang\supit{c}, 
 Fabiola Iba\~nez-Romano\supit{a}, Homin Jiang\supit{a},
Yu-Wei Liao\supit{c}, Kai-Yang Lin\supit{a}, Guo-Chin Liu\supit{a}, 
Sandor Molnar\supit{a}, 
Hiroaki Nishioka\supit{a}, Keiichi Umetsu\supit{a}, Fu-Cheng Wang\supit{c},
Jiun-Huei Proty Wu\supit{c},
Pablo Altamirano\supit{d}, Chia-Hao Chang\supit{a}, Shu-Hao Chang\supit{a}, 
Su-Wei Chang\supit{a}, Chi-Chiang Han\supit{a},  
Derek Kubo\supit{d}, Chao-Te Li\supit{a}, Pierre Martin-Cocher\supit{a}, Peter Oshiro\supit{d}
\skiplinehalf
\supit{a}Academia Sinica, Institute of Astronomy and Astrophysics,
                 P.O.Box 23-141, Taipei 10617, Taiwan \\
\supit{b}Australia Telescope National Facility, P.O.Box 76, Epping NSW 1710, Australia\\
\supit{c}Department of Physics, National Taiwan University,
No.1 Sec.4 Roosevelt Road, Taipei 10617, Taiwan\\
\supit{d} ASIAA Hawaii Site Office, 645 N.A'ohoku Place, Hilo, Hawaii 96720
}

\authorinfo{Further author information: \\
E-mail: pmkoch@asiaa.sinica.edu.tw, Telephone: 00886 2 33 65 22 00 ext.758}

\begin{document} 
 \maketitle




\begin{abstract}

The Array for Microwave Background Anisotropy (AMiBA) is a radio 
interferometer for research in cosmology, currently operating 7 0.6m diameter 
antennas 
co-mounted on a 6m diameter platform driven by a hexapod mount.
AMiBA is currently the largest hexapod telescope. 
We briefly summarize the hexapod operation with the current pointing error model.
We then focus on the upcoming 13-element expansion with its potential difficulties
and solutions. 
Photogrammetry 
measurements of the platform reveal deformations at a level which can affect the 
optical pointing and the receiver radio phase. In order to prepare for the 13-element
upgrade, 
two optical telescopes are installed on the platform to correlate optical pointing 
tests. Being mounted on different locations, the residuals of the two sets of 
pointing errors show a characteristic phase and amplitude difference as a function 
of the platform 
deformation pattern. These results depend on the telescope's azimuth, elevation and polarization
position. An analytical model for the deformation is derived in order to 
separate the local deformation induced error from 
the real hexapod pointing error. Similarly, we demonstrate that the 
deformation induced radio phase error can be reliably modeled and calibrated, 
which allows 
us to recover the ideal synthesized beam in amplitude and shape of 
up to 90\% or more. 
The resulting array efficiency and its limits are discussed based on the derived errors.

\end{abstract}

\section{Introduction}
 
\begin{figure}
\begin{center}
\includegraphics[scale=0.95]{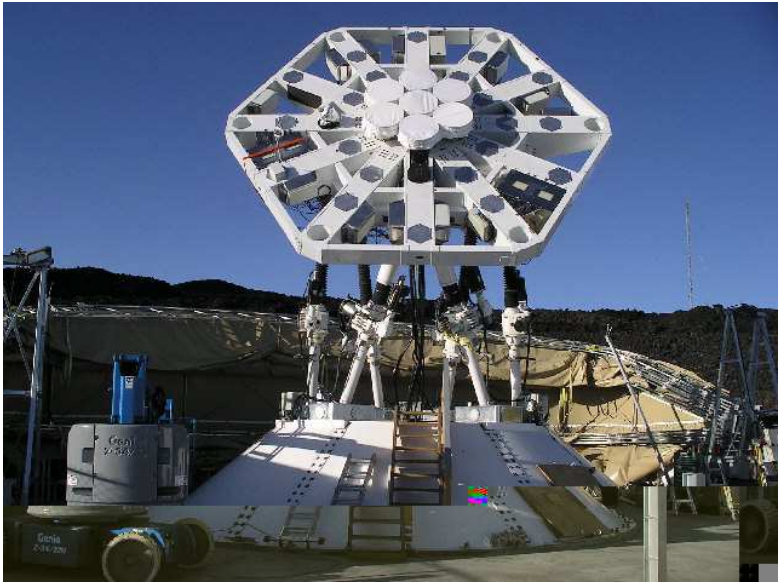}
\caption{\label{front_hexpol0_bw} Front view of the AMiBA.
Installed are 7 antennas in compact configuration, giving 0.6m, 1.04m and 1.2m baselines.
Free receiver holes in the platform allow for different array configurations and for the 
expansion phase with 13 antennas. The 1st 8 inch refractor (OT1) for optical pointing is attached to  
the black bracket below the lowermost antenna at a distance of 1.4m from the platform center.
A second optical telescope (OT2) is located at 90$^{\circ}$ to the left hand side at the same radius.}
\end{center}
\end{figure}

The Array for Microwave Background Anisotropy (AMiBA) is a forefront radio interferometer 
for research in cosmology. The project is led by the Academia Sinica, 
Institute of Astronomy and Astrophysics (ASIAA), Taiwan, with major collaborations 
with National Taiwan University, Physics Department (NTUP), Electrical Engineering 
Department (NTUEE), and the Australian Telescope National Facility (ATNF). 
Contributions also came from the Carnegie Mellon University and NRAO. 
As a dual-channel 86-102 GHz interferometer array of up to 19 elements, 
AMiBA is designed to have full polarization capabilities, sampling structures 
greater than $2^{\prime}$ in size. The AMiBA targets the distribution of clusters of 
galaxies via the Sunyaev-Zel'dovich (SZ) Effect$^1$. It will also measure the 
Cosmic Microwave Background (CMB) temperature anisotropies$^2$ on scales 
which are sensitive to structure formation scenarios of the Universe. 
AMiBA is sited on Mauna Loa, Big Island, Hawaii, at an elevation of 3,425m, 
currently operating the initial phase with 7 antennas 
of 0.6m diameter in a compact configuration, 
Fig.\ref{front_hexpol0_bw}.  
The correlator and the receivers
with the antennas are co-mounted on a fully steerable 6m diameter platform controlled by 
a hexapod mount. 
The AMiBA hexapod mount with its local control system was designed and manufactured by 
Vertex Antennentechnik GmbH, Duisburg, Germany. 
The 0.6m diameter Cassegrain antenna$^3$ field-of-view (fov) is $\sim 22^{\prime}$ at Full Width Half 
Maximum (FWHM). The synthesized beam resolution in the hexagonally 
compact configuration (0.6m, 1.04m and 1.2m baselines) is $\sim 6^{\prime}$.
This sets our target pointing accuracy to about $\sim 0.6^{\prime}$. \\
In parallel to the current 7-element observation program, AMiBA is undergoing a 13-element 
expansion with 1.2m diameter Cassegrain antennas (fov $\sim 11^{\prime}$). The resulting synthesized
beam of about $\sim 2^{\prime}$ requires then ideally an improved pointing accuracy of about $\sim 0.2^{\prime}$.
The platform photogrammetry tests$^{4,5}$ have revealed local deformations
which might affect individual antenna pointings (including the optical telescope pointing)
 and their radio phases for the longer baselines.
It therefore seems desirable to include such a correction in the current pointing error model
and in the phase correction scheme. \\  
The goal of this paper is to investigate a refined pointing error and phase correction
 model for the AMiBA expansion 
phase.
The adopted  approach can be of general interest 
since a platform deformation for a hexapod of this size is likely to be a generic problem
depending on the level of accuracy which is needed. 
The paper is organized as following: Section \ref{7element} summarizes the pointing error model
used for the currently operating 7-element compact configuration. Section \ref{13element}
presents the approach taken to improve the pointing accuracy 
and the phase correction for the 13-element expansion
and it also discusses its limitations.
Previous AMiBA status reports and the current observations are described in a series of papers.$^{6,7,8,9}$

\begin{figure}
\begin{center}
\includegraphics[scale=0.8]{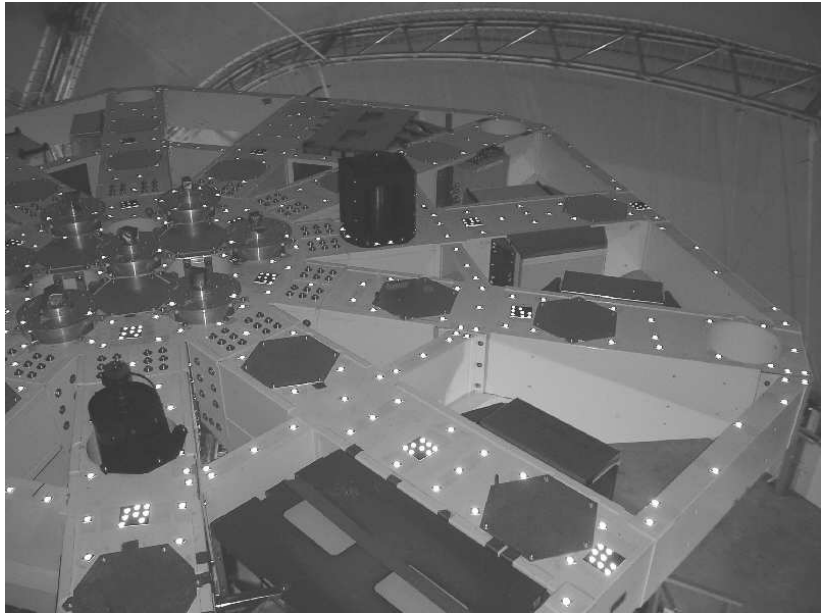}
\caption{\label{ot1_ot2__bw}OT1 (in the back) and OT2 (in the front, lower left hand side corner),
mounted with a $90^{\circ}$ phase shift in between them at a radius of 1.4m. 
Clearly seen are also the reflecting photogrammetry targets.}
\end{center}
\end{figure}

\section{7-element pointing} \label{7element}

\begin{center}
\begin{figure}
\begin{center}
\includegraphics[scale=0.65]{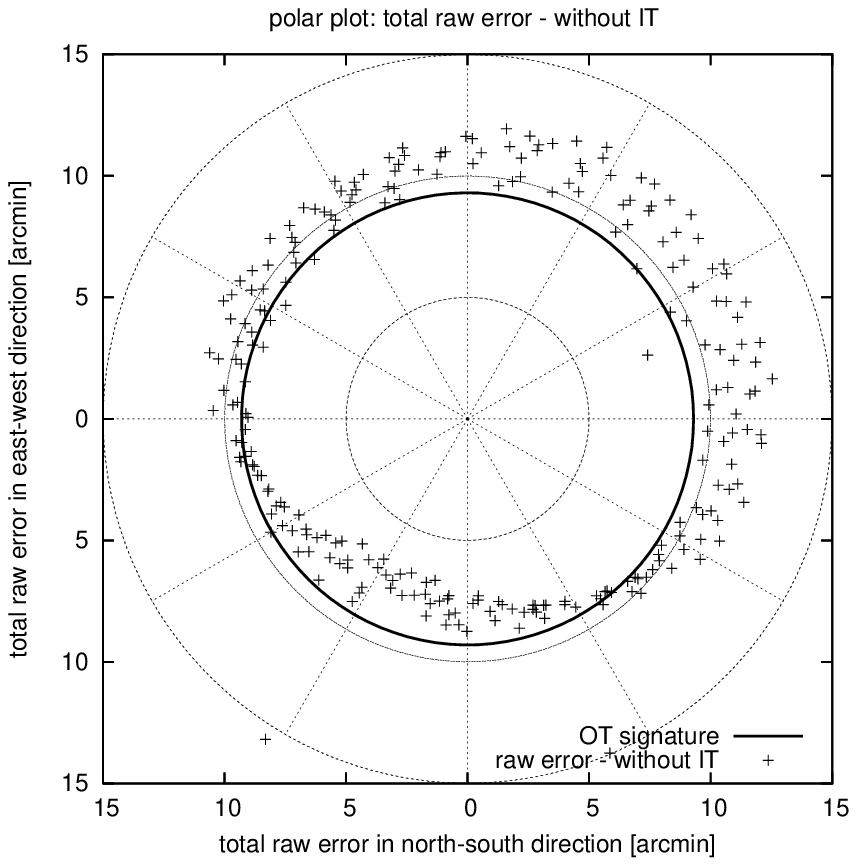}
\includegraphics[scale=0.65]{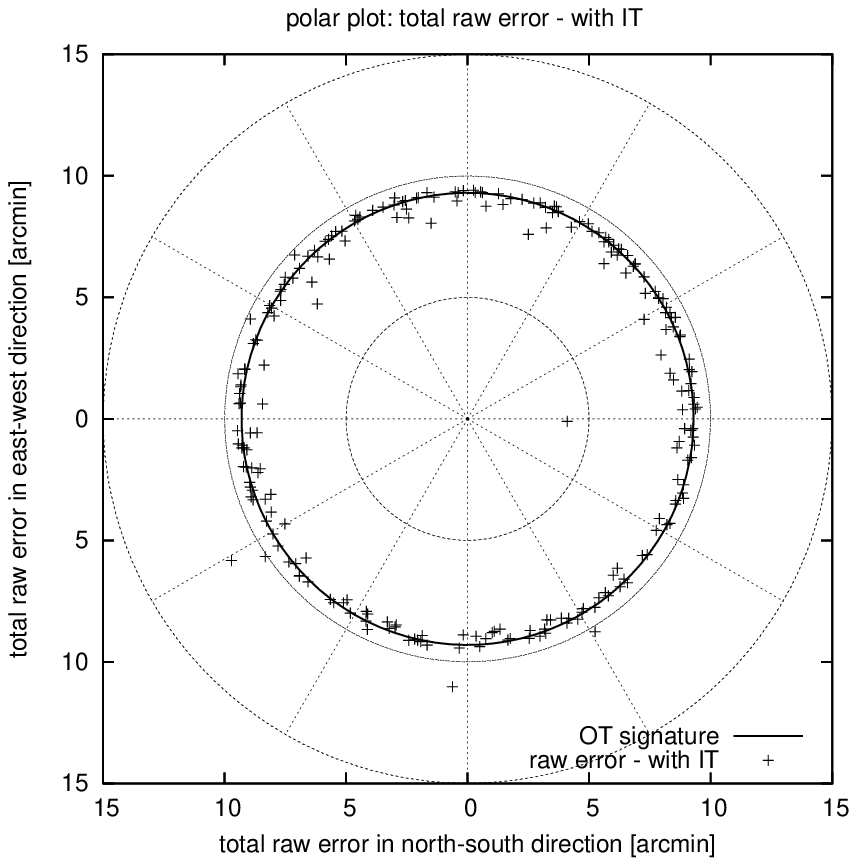}
\caption{\label{it} Left hand side figure: 
Polar plot of the total pointing error (azimuth and elevation component combined)
without IT correction. The  OT collimation error is shown as a 
circle which represents the OT tilt $\sim 9^{\prime}$ with respect to the mount pointing axis. 
The data points scattered off the circle show the residual pointing errors 
 for the IT, leading to $\sim 1^{\prime}$ rms pointing error. 
The test was done with 250 stars.
Right hand side figure: Polar plot of the total pointing error with the IT correction. The reduced 
scatter verifies the improved pointing with $\sim 0.4'$ rms pointing error.}
\end{center}
\end{figure}
\end{center}

The AMiBA hexapod has an elevation limit of $30^{\circ}$ with a continuous azimuth range and a polarization
limit of $\pm30^{\circ}$ at each azimuth-elevation position. 
An important consequence of the hexapod mount are the non-existent azimuth and elevation encoders 
compared to more traditional telescopes.  The 3-dimensional locations of the upper and lower 
universal joints in the reference positions and the 6 exact encoder-measured jack lengths  
completely define the geometry of the mount. The 6 upper universal joint 
locations define a best-fit plane with its normal defining the resulting pointing axis.
During the operation it is thus most essential to control the lengths of the 6 jacks.
In order to achieve a pointing accuracy of $\sim 0.6^{\prime}$, a 2-step approach$^4$ was adopted:
\begin{itemize}
\item
first pointing run: \\
all known pointing corrections are activated.  
The star position on the  optical telescope  (OT) ccd is split into an azimuth and an elevation error as a 
function of the mount position.  The errors are analyzed to separate the OT collimation error from the  
mount pointing error.
The resulting residual errors are filled in a 3-dimensional  error interpolation table (IT) as a 
function of elevation, azimuth and polarization.
\item
second pointing run:\\
 all known pointing corrections and the IT 
are activated and the  improvement is verified.
\end{itemize}
The pointing error model can be listed in two groups:\\
\begin{itemize}
\item
Jack corrections: \\
- {\it jack pitch error compensation:}\\
calibrated at reference temperature, look-up table as a function of jack length for each jack\\
 - {\it jack temperature compensation:}\\
analytical model, real-time measurement with 3 temperature sensors along the jacks\\
				- {\it jack rotation correction:} \\
analytical model for encoder undetected jack rotation\\
				- {\it support cone compensation mode:}\\
model for universal joint shifts due to temperature changes, 
monitored with several temperature sensors
\item
Telescope corrections: \\
- {\it radio/optical refraction:} \\
analytical refraction model, wavelength dependent\\ 
- {\it OT correction (collimation error):}\\
analytical model, collimation correction derived from pointing data\\
				- {\it anchor cone orientation correction:}\\
analytical model, rotation of mount with respect to north\\
				- {\it error interpolation table (IT):}\\
interpolation table, residual errors which are not explicitly modeled, derived from pointing data\\
\end{itemize}
The 2-step approach with the above outlined pointing error model and the additional improvement 
from the IT is illustrated in Fig.\ref{it}.
The original pointing error from $\sim 1^{\prime}$ rms is reduced to $\sim 0.4^{\prime}$ rms with the help of the 
IT.
These first results were derived with a single OT (OT1) on the platform (Fig.\ref{ot1_ot2__bw}).

\section{13-element pointing} \label{13element}

\subsection{Comparison OT1 - OT2}

\begin{figure} 
\begin{center}
\includegraphics[scale=0.8]{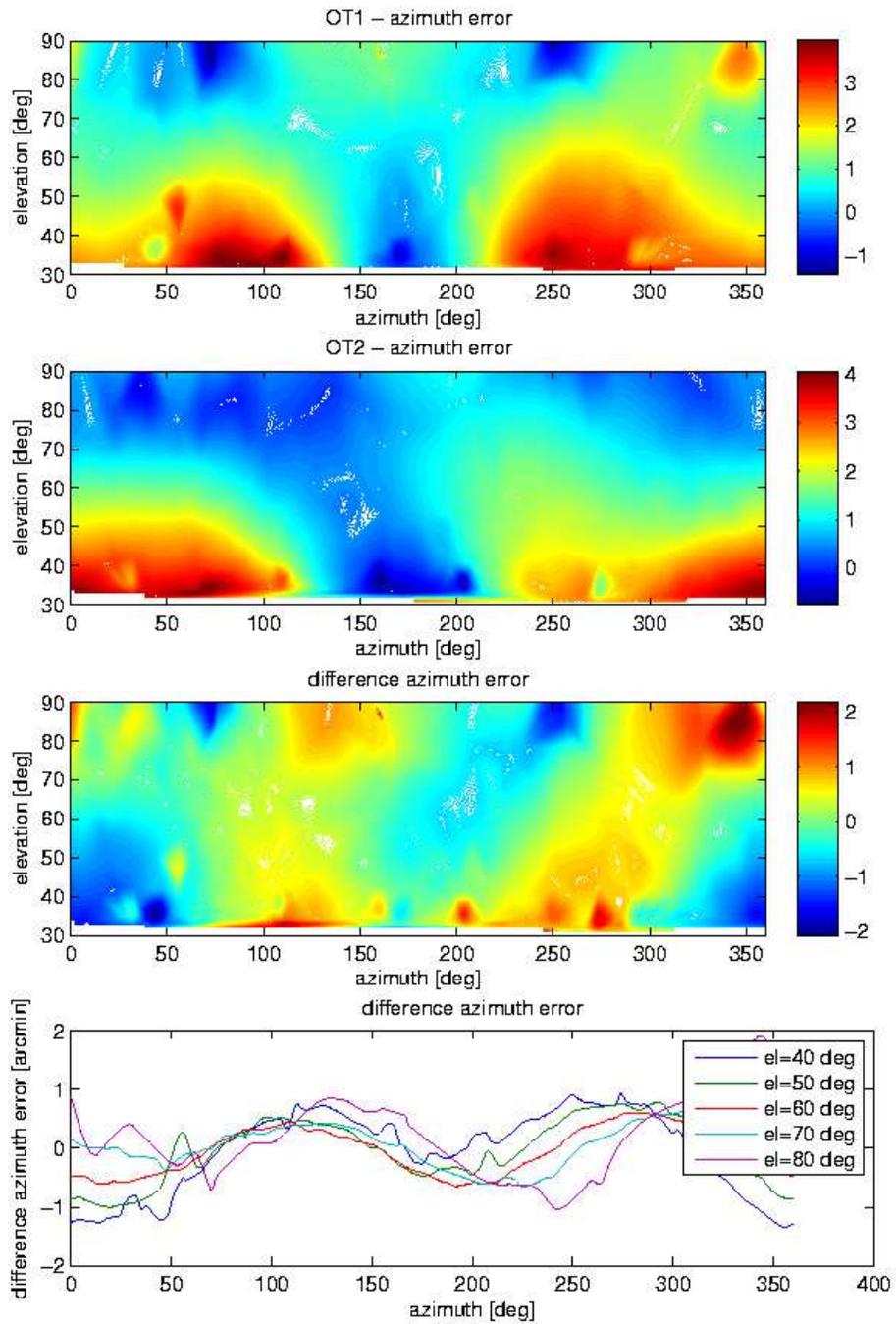}
\caption{\label{ot12_az_combined}Upper two panels: OT1 and OT2 azimuth errors after 
removing the OTs collimation errors. Both OTs show similar dominating structures.
Lower two panels: difference in azimuth errors shown in a contour plot and extracted 
for some elevation bands. A clear oscillating structure with positive and negative
domains becomes visible. All units are in arcmin.  }
\end{center}
\end{figure}

\begin{figure}
\begin{center}
\includegraphics[scale=0.8]{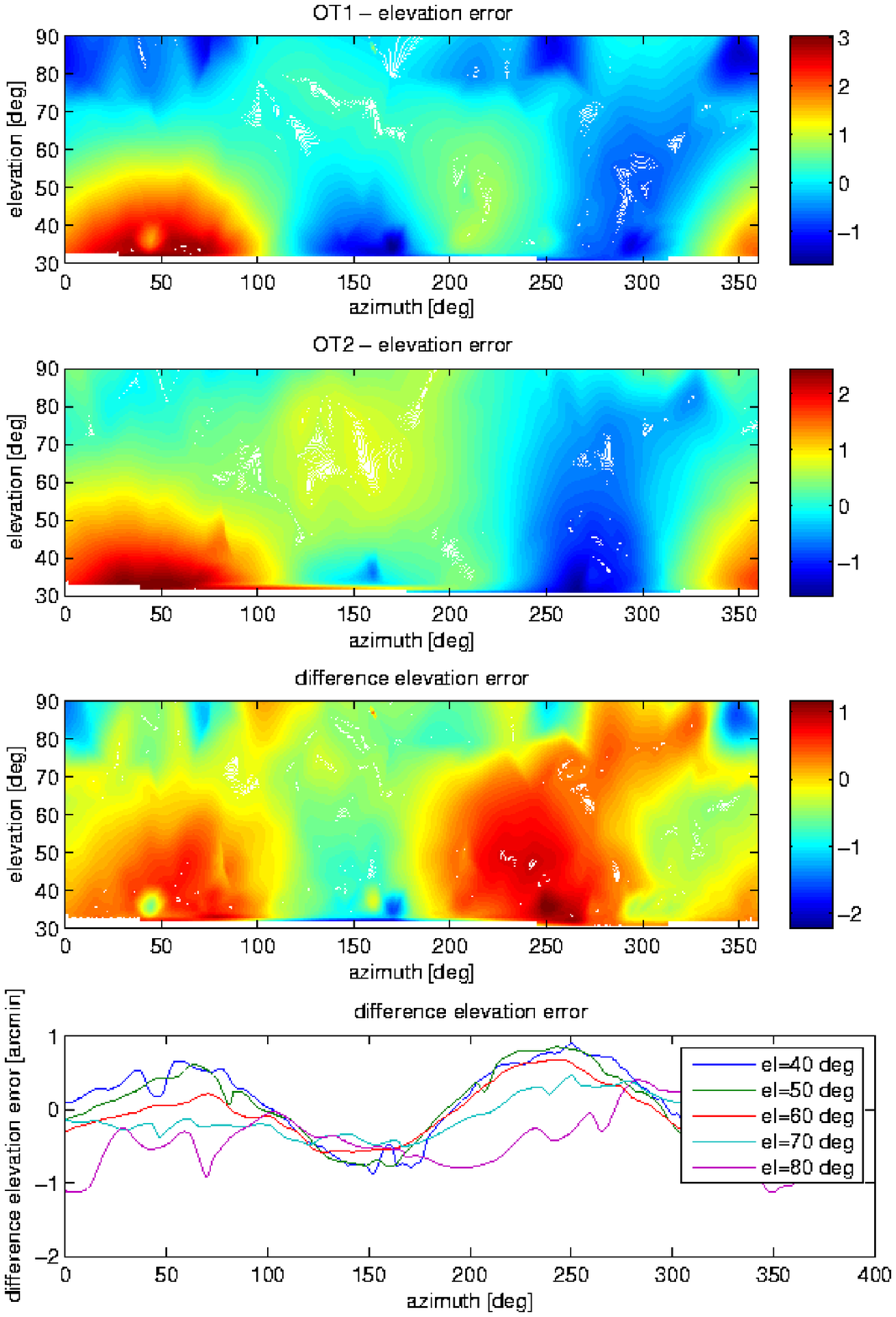}
\caption{\label{ot12_el_combined}As in Fig.\ref{ot12_az_combined} but for the elevation error. 
Again, a similar structure in the OT1 and OT2 data is apparent, as well as the oscillating
feature.}
\end{center}
\end{figure}

The mount pointing can only account for an average pointing. Any possible local irregularity or 
(platform) deformation is averaged out and therefore not taken properly into account.
This is in particular also the case for each antenna and the OTs.
Therefore, the results in section \ref{7element} have to be re-analyzed for an improved pointing.
A second optical telescope (OT2) was installed on the platform in order to cross-check the 
pointing results from OT1. This also provides an independent and complementary  
tool to the photogrammetry tests. 
The OT2 (SBIG ST-237A ccd with a hyperstar f/2.0. lens, fov $\sim 42'$ x $32'$) 
is installed at an angle of $90^{\circ}$ with 
respect to the OT1 at the same radius r=1.4m from the platform center, Fig.\ref{ot1_ot2__bw}. 
It is also integrated
in an automatic pointing schedule$^4$ which allows us to simultaneously 
take ccd images with both OTs. \\
The Figs.\ref{ot12_az_combined} and \ref{ot12_el_combined} show the separated azimuth and 
elevation errors for both OT1 and OT2 after the OTs collimation errors
(solid circle in Fig.\ref{it}) have been removed. 
The pointing run was done with 200 stars, equally distributed with constant
solid angle over the total accessible sky. 
The platform polarization was kept at zero for this test.
The errors are then interpolated (cubic interpolation) on a regular 
azimuth-elevation grid. 
Apparent similar pointing errors are measured by both OT1 and OT2, 
Figs.\ref{ot12_az_combined} and \ref{ot12_el_combined} upper two panels, for both azimuth and 
elevation errors with some clear dominating patches
at lower elevation.
The overall azimuth and elevation rms errors between OT1 and OT2 agree within $4-7''$. \\
However, taking the difference between OT1 and OT2 for their corresponding azimuth and 
elevation errors reveals clear systematic patterns 
(Figs.\ref{ot12_az_combined} and \ref{ot12_el_combined} lower two panels) with an amplitude of 
up to $1^{\prime}$.

In the ideal case, this difference should be negligible. If there is a local error contribution
at each OT location, this measures the difference between the local errors, symbolically written
as:
\begin{eqnarray*}
error_{OT1}=local_{OT1} + pntgerror \nonumber \\
error_{OT2}=local_{OT2} + pntgerror  \nonumber \\
difference=local_{OT1}-local_{OT2} \nonumber \\
\end{eqnarray*}
The difference plots show clear 'plus' and 'minus' patches, which are likely the result
of the platform deformation (section \ref{platform})
which sometimes tilts the OTs towards each other and sometimes
away from each other. Over a full azimuth circle the OTs are aligned with each other four times
(2 local maxima and 2 local minima on the saddle pattern, Fig.\ref{deformation_ot}) and the difference becomes zero.
The bottom panels in the Figs.\ref{ot12_az_combined} and \ref{ot12_el_combined} illustrate 
this for some elevation bands.

\subsection{Platform deformation modeling}  \label{platform}

\subsubsection{Optical pointing}

\begin{figure}
\begin{center}
\includegraphics[scale=0.9]{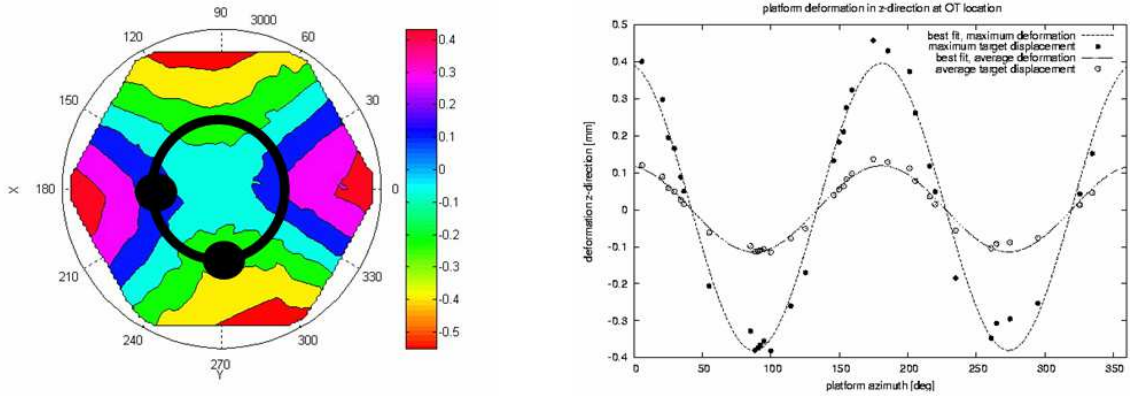}
\caption{\label{deformation_ot}
Left hand side figure: A typical saddle-type platform deformation in z-direction, measured
from photogrammetry data. The units are in mm. 
The two black dots along the black circle mark the locations of the two OTs. 
Right hand side figure:
Platform deformation in z-direction as a function of platform azimuth 
at a radius r=1.4m where the OTs are mounted. The selected data points are within an annulus
of r=1.4m$\pm0.2m$, extracted along the black circle in the left hand side figure.  The maximum deformation
is measured at the extreme position (az,el,pol)=(0,30,20) and sets therefore an upper limit to 
the deformation error of about $1^{\prime}$. 
Also illustrated for comparison is an average mount position (az,el,pol)=(0,60,00) with an 
amplitude of about 0.11 mm, leading to a corresponding OT tilt of about $0.25^{\prime}$.   
Clearly visible is the saddle structure with a functional form $\cos(2az)$. 
}
\end{center}
\end{figure}

Since the mount pointing can only correct for an average pointing error, any 
local deformation error should be separated.
In order to do that, the photogrammetry
results$^4$ are used to construct an analytical model for the local platform deformation 
induced pointing error. 
The saddle-type deformation in z-direction 
at a fixed telescope position can be modeled fairly well with a function of type 
$f(x,y)=\bar{A}(x^2-y^2)$,
both for the entire
platform and for an annulus at a radius $r$ (Fig.\ref{deformation_ot}).
The OT at any location has its optical pointing axes normal to the tangential plane at the deforming platform.
The function $f$ describing the deformation with its gradient 
$\nabla f(r,\alpha)=(\frac{\partial f}{\partial r},\frac{1}{r}\frac{\partial f}{\partial\alpha})$
in spherical coordinates $x=r\cos\alpha, y=r\sin\alpha$ are written as:
\begin{eqnarray}
f(r,\alpha)&=& \bar{A}r^2(\cos^2\alpha-\sin^2\alpha),\\
\frac{\partial f}{\partial r}&=& 2\bar{A}r(\cos^2\alpha-\sin^2\alpha),               \label{tangential_vector_1}\\
\frac{1}{r}\frac{\partial f}{\partial\alpha}&=& -4\bar{A}r\cos\alpha\sin\alpha.         \label{tangential_vector_2}
\end{eqnarray}

In order to calculate the azimuth and elevation error components for the induced pointing error, the 
angles $d$ between the normal of the tangential plane ($\hat{n}$) and the normal to the platform 
$\hat{n_0}$ along the azimuthal
and radial direction are calculated as a function of the platform azimuth $\alpha$ and radius $r$:

\begin{equation}
d=\angle(\hat{n_0},\hat{n})=\pi/2 - \angle(\hat{n_0},\hat{t}),
\end{equation}
with $\hat{t}$  being either one of the unit tangential vectors derived in Eqs.(\ref{tangential_vector_1}) 
and (\ref{tangential_vector_2}). 
Using the scalar product $\cos d(\hat{n_0},\hat{n})=\hat{n_0}\cdot\hat{n}$, we find the 
deformation induced pointing errors in azimuthal and radial
direction as:

\begin{eqnarray}
d_{az}&=&\pi/2 - \arccos\left(\frac{-4R\bar{A}\cos\alpha\sin\alpha}{\sqrt{1+16R^2\bar{A}^2\cos^2\alpha\sin^2\alpha}} \right),  
\label{az_error}\\
d_{el}&=&\pi/2 - \arccos\left(\frac{2R\bar{A}(\cos^2\alpha-\sin^2\alpha)}{\sqrt{1+4R^2\bar{A}^2(\cos^2\alpha-\sin^2\alpha)}}\right),
\label{el_error}  
\end{eqnarray}
where $r\equiv R$ is the radius where the OTs are mounted. 
Having two OTs mounted on the platform at the same radius $R$ with a separation of 90$^{\circ}$ in between them
 (Fig.\ref{ot1_ot2__bw})
provides for each of the error components in Eqs.(\ref{az_error}) and (\ref{el_error}) a set of two equations with a 
characteristic 90$^{\circ}$ phase shift in $\alpha$.
Additionally, the saddle pattern has an offset $\phi$ relating the saddle pattern orientation to the mount 
position, i.e. $\alpha=az+\phi$ for OT1 and $\alpha=az+\phi+\pi/2$ for OT2. The deformation is then 
entirely described with the functional form in the Eqs.(\ref{az_error}) and (\ref{el_error}) and the 
two fitting parameters $\bar{A}$ and $\phi$.\\
We remark that the Eqs.(\ref{az_error}) and (\ref{el_error}) are almost identical to the function of type
$A\cos(2az+\phi)$. For simplicity we use the later one in section \ref{correlation} where 
$\bar{A}$ and $R$ are absorbed in the amplitude $A$.


\subsubsection{Radio pointing: radio phase correction and synthesized beam}

\begin{center}
\begin{figure}
\includegraphics[scale=0.55]{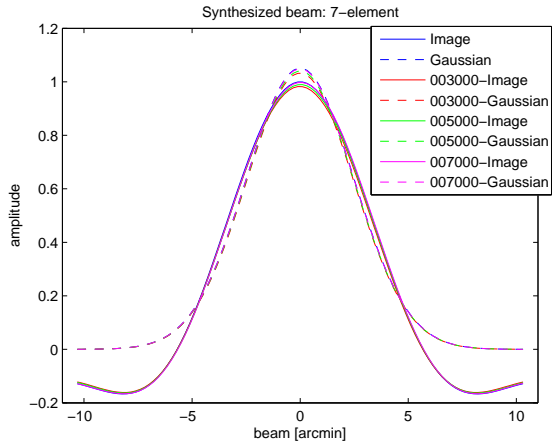}
\includegraphics[scale=0.55]{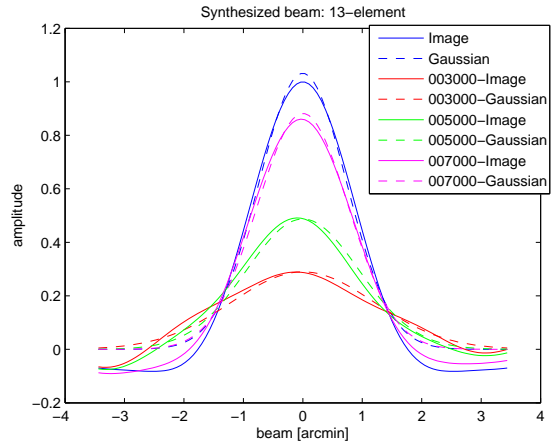}
\includegraphics[scale=0.59]{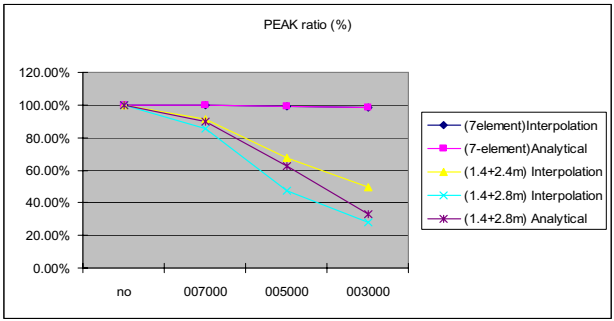}
\includegraphics[scale=0.6]{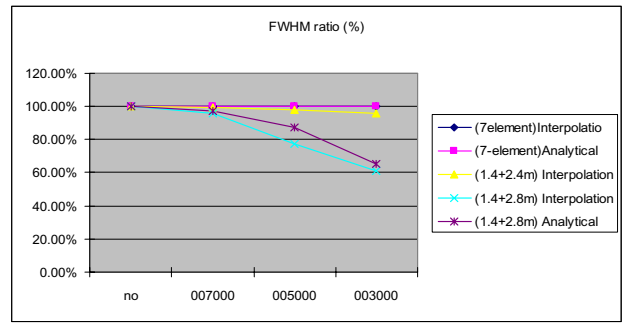}
\caption{\label{beam_ratio}Top panels: The distorted synthesized beams (labeled as 'Image') 
for the 7-element and 13-element 
configuration with their best-fit Gaussian beams. Different curves are labeled with the mount position in 
azimuth-elevation-polarization,
with the first curve being a reference position without deformation. An 'Image' is calculated from the 
measured phase errors and represents therefore a point source image. 
 For the current 7-element compact configuration (left hand side panel) the 
distortion is negligible.
Bottom panels: Gaussian beam peak and FWHM ratios (with respect to an ideal no-deformation case)
 for different configurations at 
different mount positions. The ratios decrease with lower elevation.
1.4+2.4m/1.4+2.8m are different possible configurations for 13-element.}
\end{figure}
\end{center}

Besides the OTs, the platform deformation also affects each individual radio antenna: the tilting of 
each antenna leads to a changing misalignment error and therefore reduces the 
efficiency (section \ref{efficiency}), the deformation correction in z-direction gives a radio phase error. 
Like the OTs measure directly the relevant quantity for the pointing error analysis (error in arcmin), 
the photogrammetry directly gives the relative phase errors (in mm) with respect to a reference position
for all antenna locations. 
Fig.\ref{beam_ratio} shows the distorted synthesized beams for a set of mount positions with 
their measured radio phase errors. For the largest deformations (up to $800\mu m$ rms), the beam 
for a 13-element configuration is 
severely distorted. Since the platform deformations are repeatable (within $\sim 50\mu m$), the deformation
can be modeled. 
Based on a set of about 50 measured mount positions, 
we have developed both an analytical model\footnote{
The model is again based on the functional form $A\cos(2az+\phi)$, with the amplitude $A$ and the 
phase $\phi$ being complicated functions of azimuth, elevation and polarization. 
}
and a 3-dimensional interpolation table approach in order to predict and correct the phase error. 
The deformation rms residuals after correction are less then $100\mu m$, which restores the synthesized
beam amplitude to more than 90\%. For the current 7-element compact configuration the beam 
distortion is negligible.

\begin{center}
\begin{figure}
\includegraphics[scale=0.38]{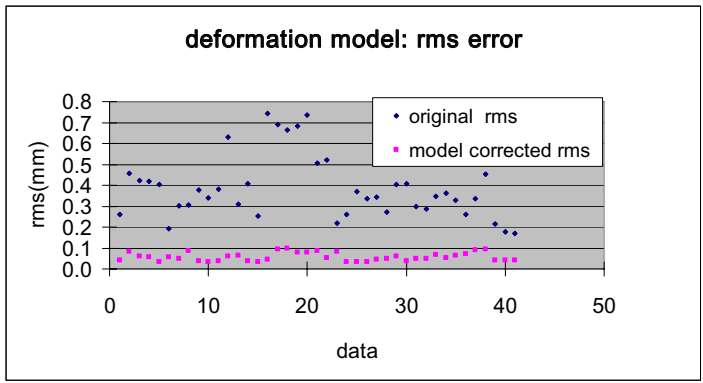}
\includegraphics[scale=0.42]{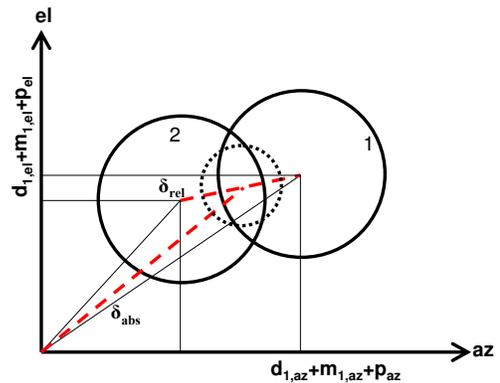}
\caption{\label{errors}Left hand side: Original platform deformation rms and residual rms after model correction
for a data set of about 40 telescope positions.
Right hand side: $\delta_{rel}$ and $\delta_{abs}$ (in long-dashed lines) 
for the efficiency calculations. The 
two antenna primary beams and the synthesized beam are shown in solid and short-dashed circles, respectively. }
\end{figure}
\end{center}

\subsection{Correlation OT1 - OT2: local platform deformation and remaining pointing error } \label{correlation}

The goal in this section is to separate the real mount pointing error from the local platform deformation
which affects the two OTs. This will then define the pointing accuracy.
In order to achieve this goal, we combine the data from OT1 and OT2, each of them providing an azimuth and 
an elevation error for each telescope position. We assume that both OTs are measuring the same underlying
pointing error ($p_{az}$ in azimuth and $p_{el}$ elevation) 
which is perturbed by the local deformation ($d_{az}$, $d_{el}$) which we describe 
with the analytical model derived in section \ref{platform}. A simultaneous fitting (non-linear curve fitting 
optimization in the least-square sense) 
is done for 4 unknowns (deformation amplitude $A$, deformation phase $\phi$, 
azimuth pointing error $p_{az}$, elevation pointing error  $p_{el}$)
with a system of 4 coupled equations\footnote{
Separating platform and pointing error at each mount position (with only 4 measurements for 
4 equations) has been found to be more reliable than trying to extract a deformation correction for 
an entire mount azimuth or elevation range. This is due to the fact that for a single specific mount 
position the deformation at the OTs locations can very well be modeled by a function of type
$A\cos(2az+\phi)$ (Fig.\ref{deformation_ot}), whereas for an azimuth or elevation range both 
phase and amplitude are complicated 
functions showing large variations (reflected in the bottom panel in the Figs.\ref{ot12_az_combined} 
and \ref{ot12_el_combined}).
}
:
\begin{eqnarray}
p^{OT1}_{az}&=& d_{az}(az,A,\phi)+p_{az},\\
p^{OT2}_{az}&=& d_{az}(az,A,\phi+\frac{\pi}{2})+p_{az},\\
p^{OT1}_{el}&=& d_{el}(az,A,\phi)+p_{el},\\
p^{OT2}_{el}&=& d_{el}(az,A,\phi+\frac{\pi}{2})+p_{el},
\end{eqnarray}
where $p^{OT1}_{az}$, $p^{OT2}_{az}$, $p^{OT1}_{el}$ and
$p^{OT2}_{el}$ are the errors measured on the CCDs after the OTs collimation errors
are removed (Figs.\ref{ot12_az_combined} and \ref{ot12_el_combined}). 
$az$ is the mount azimuth position for each set of pointing data. $d$ is the function 
describing the platform deformation, derived in section \ref{platform}. The deformation terms $d$
for OT1 and OT2 are coupled through a phase shift $\pi/2$, because the two OTs are mounted on the platform
with a $90^{\circ}$ separation in between them\footnote{
For practical purposes, we use $d_{az}=A\cos(2az+\phi)$ and $d_{el}=A\cos(2az+\phi+\pi/2)$
as approximations to the exact functions derived in section \ref{platform}.
}
.\\

\begin{figure}
\begin{center}
\includegraphics[scale=0.8]{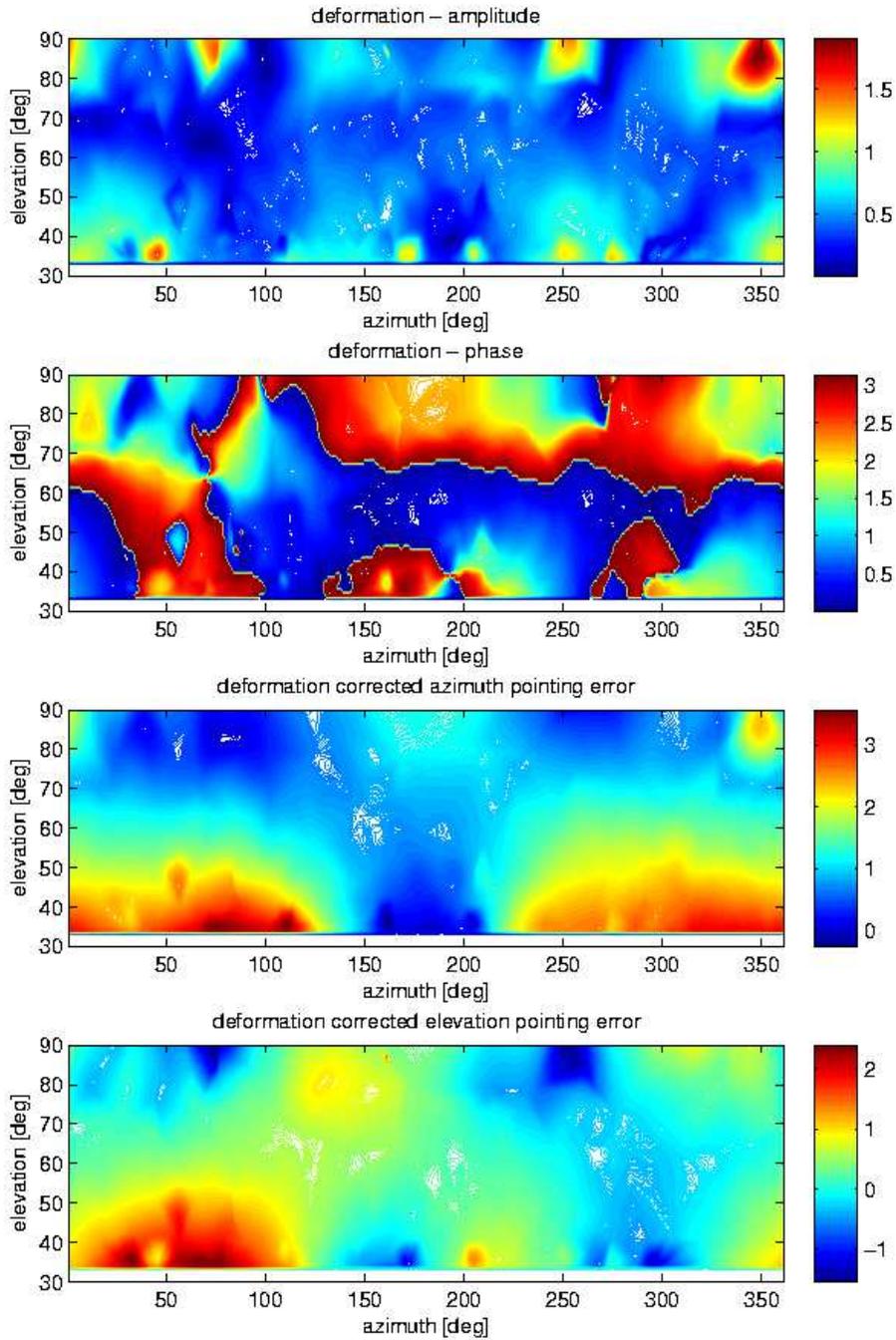}
\caption{\label{separation}The 4 fitting parameters for the simultaneous optimization for OT1 and OT2 for 
azimuth and elevation error. Units are in arcmin, except for the phase which is in rad. The lower two 
panels show the deformation corrected azimuth and elevation pointing errors $p_{az}$ and $p_{el}$,
respectively.}
\end{center}
\end{figure}

\begin{figure}
\begin{center}
\includegraphics[scale=0.39]{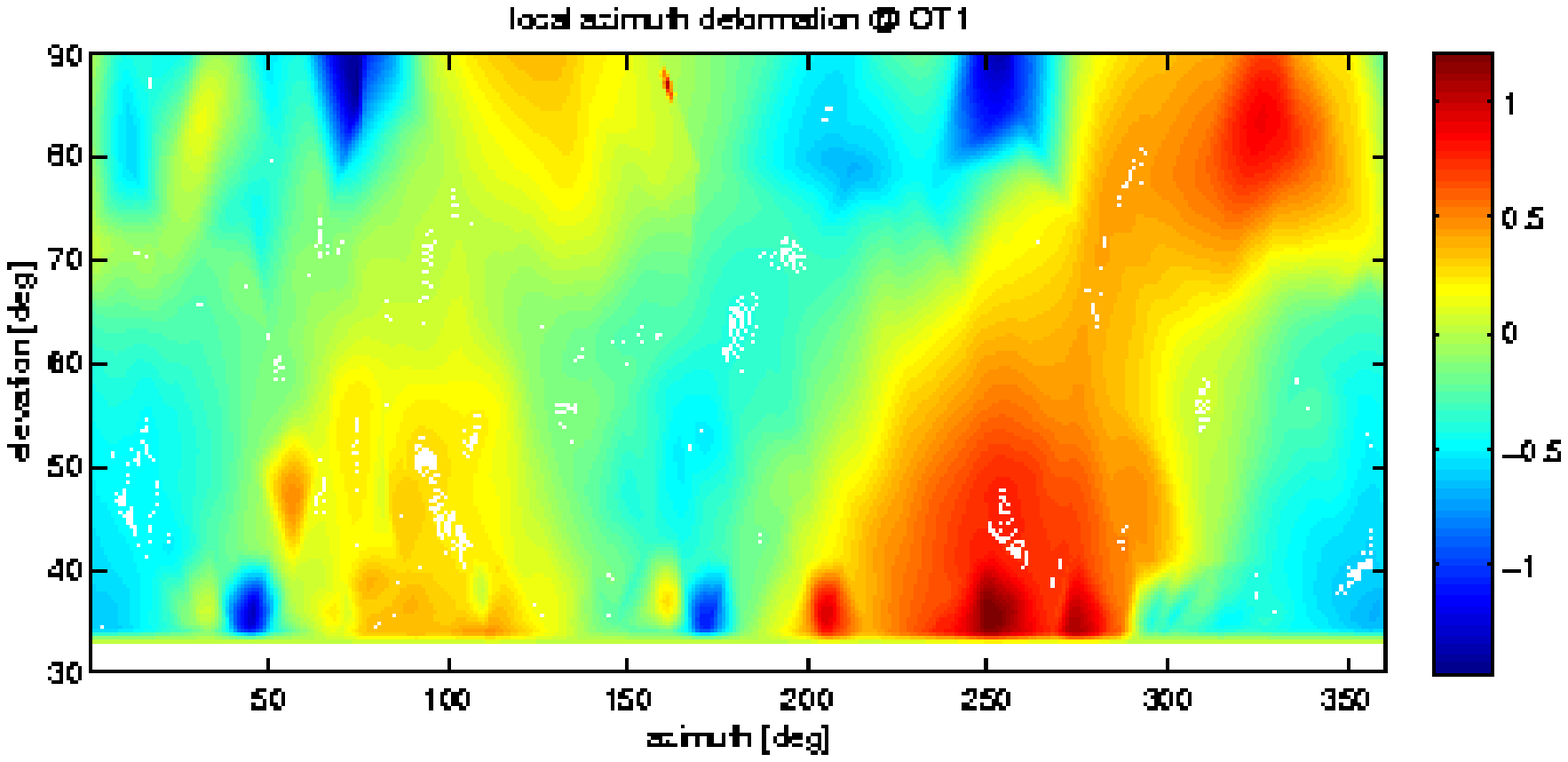}
\includegraphics[scale=0.39]{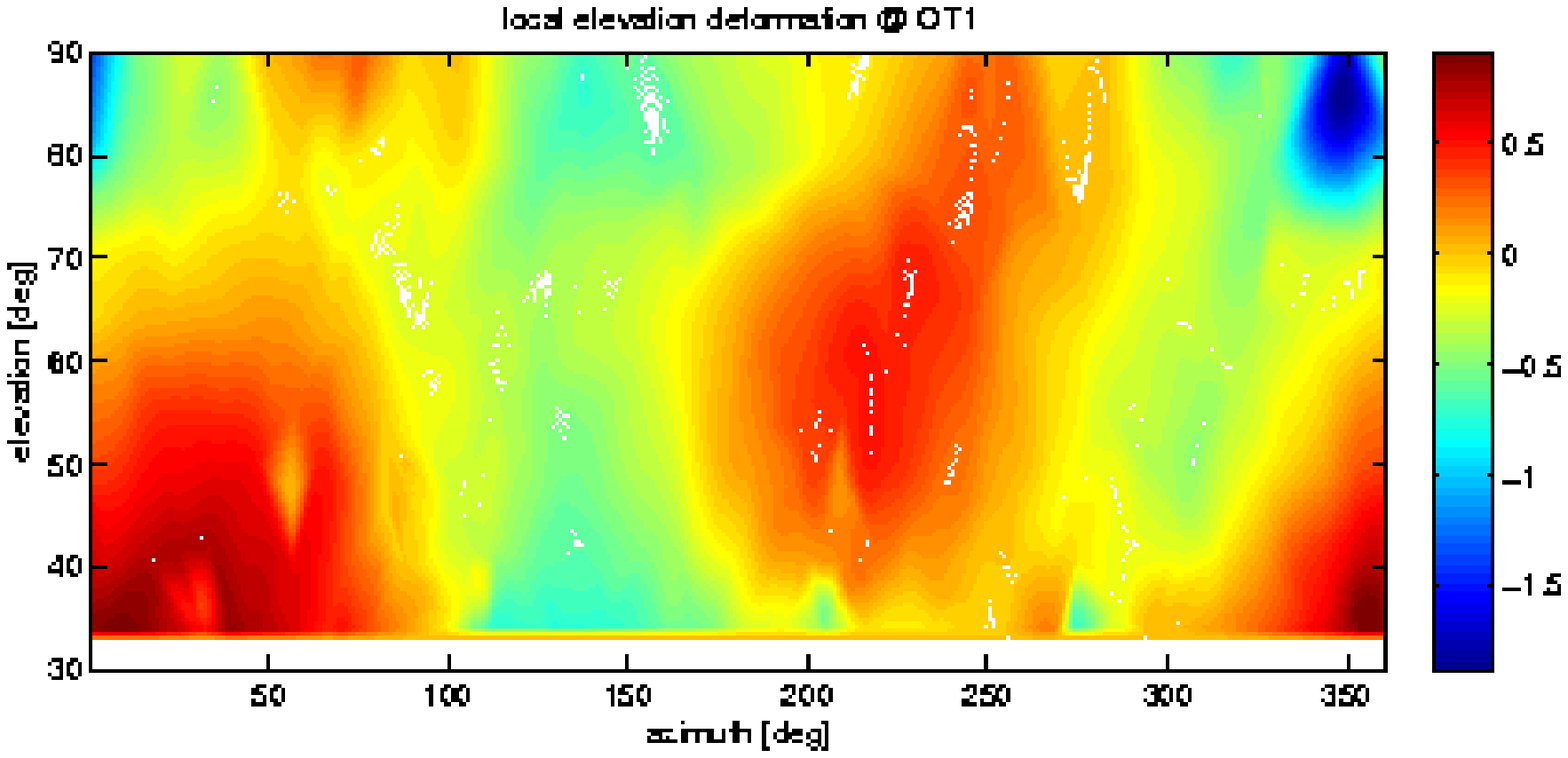}
\includegraphics[scale=0.39]{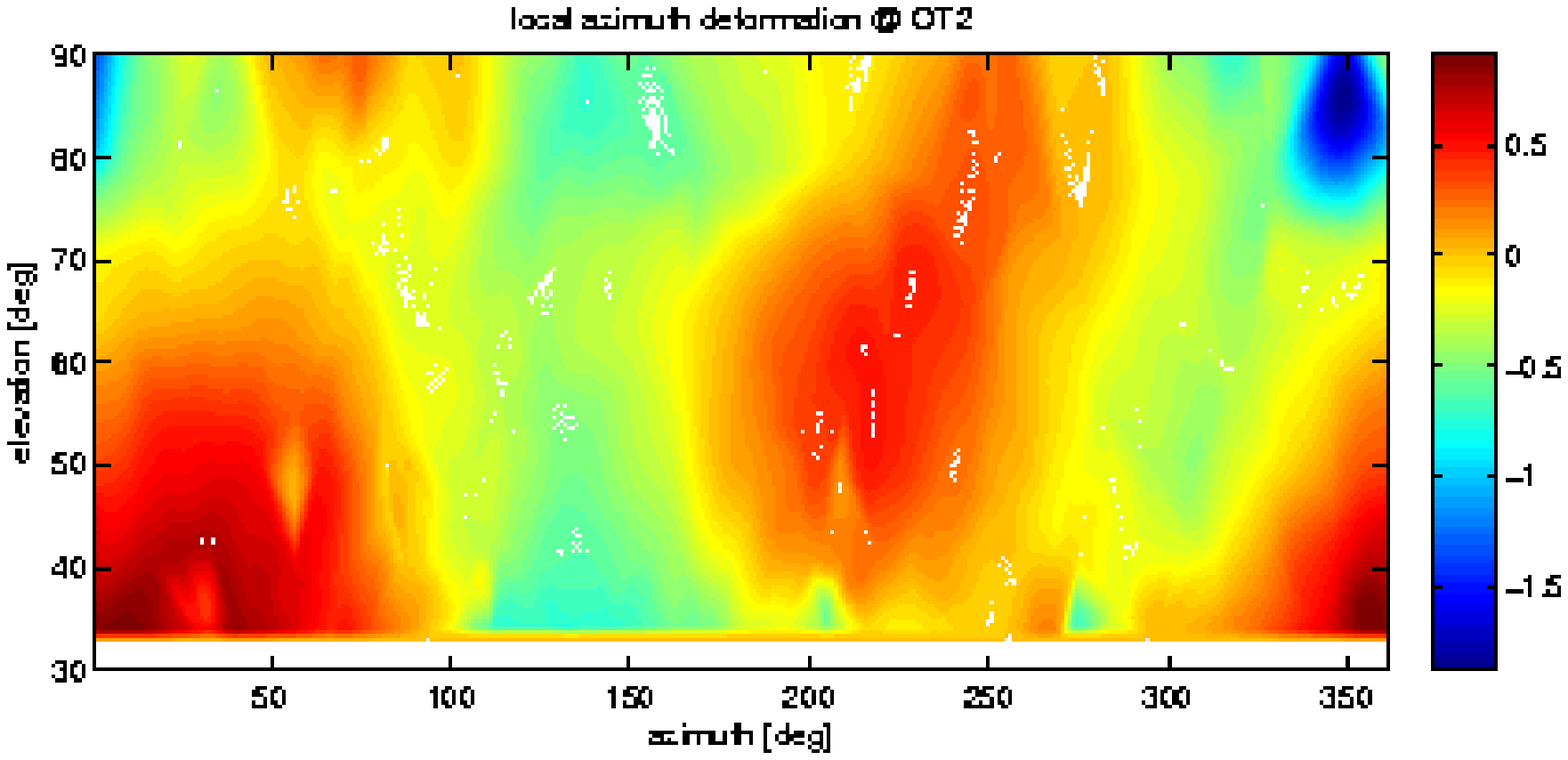}
\includegraphics[scale=0.39]{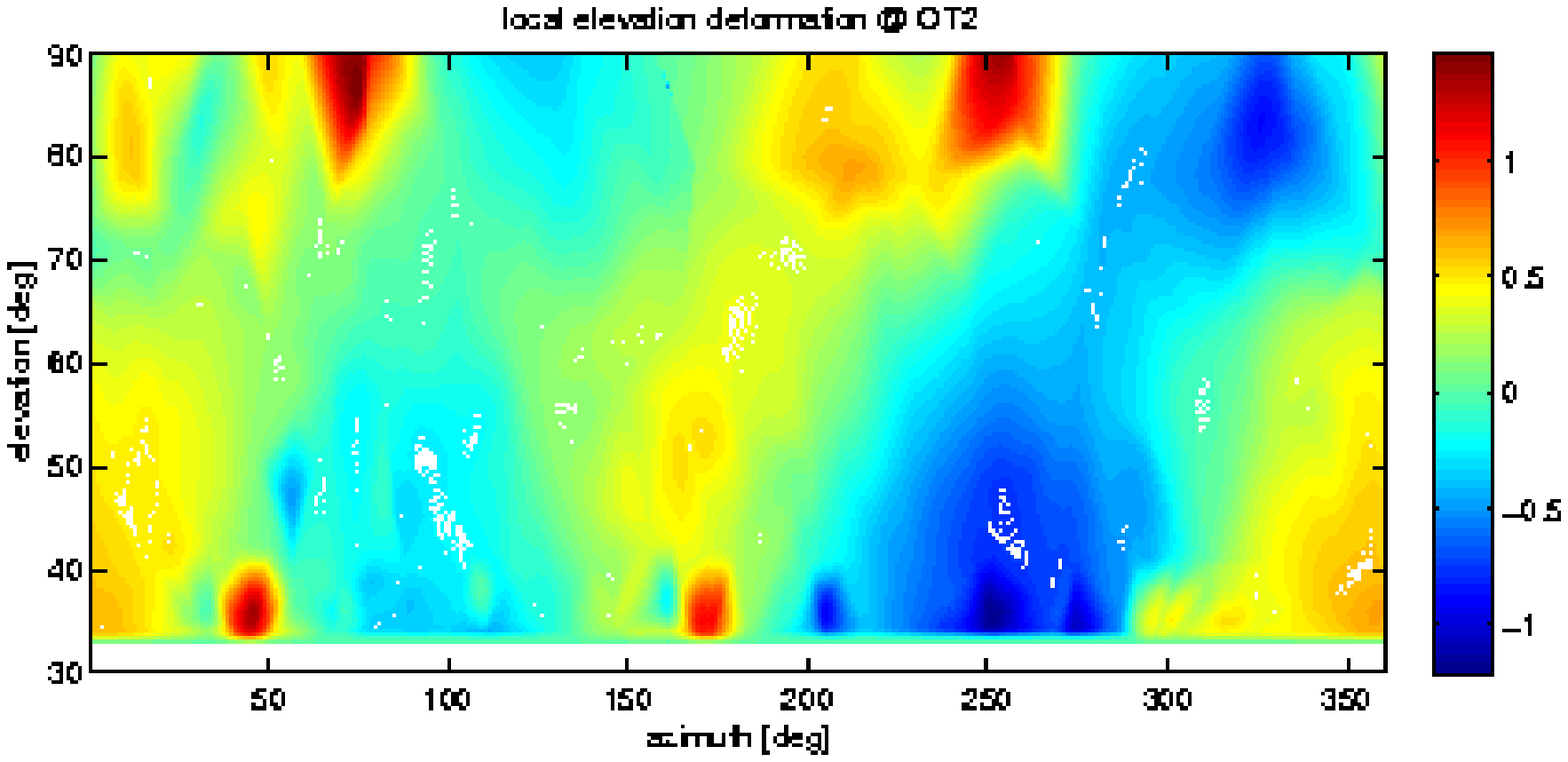}
\includegraphics[scale=0.39]{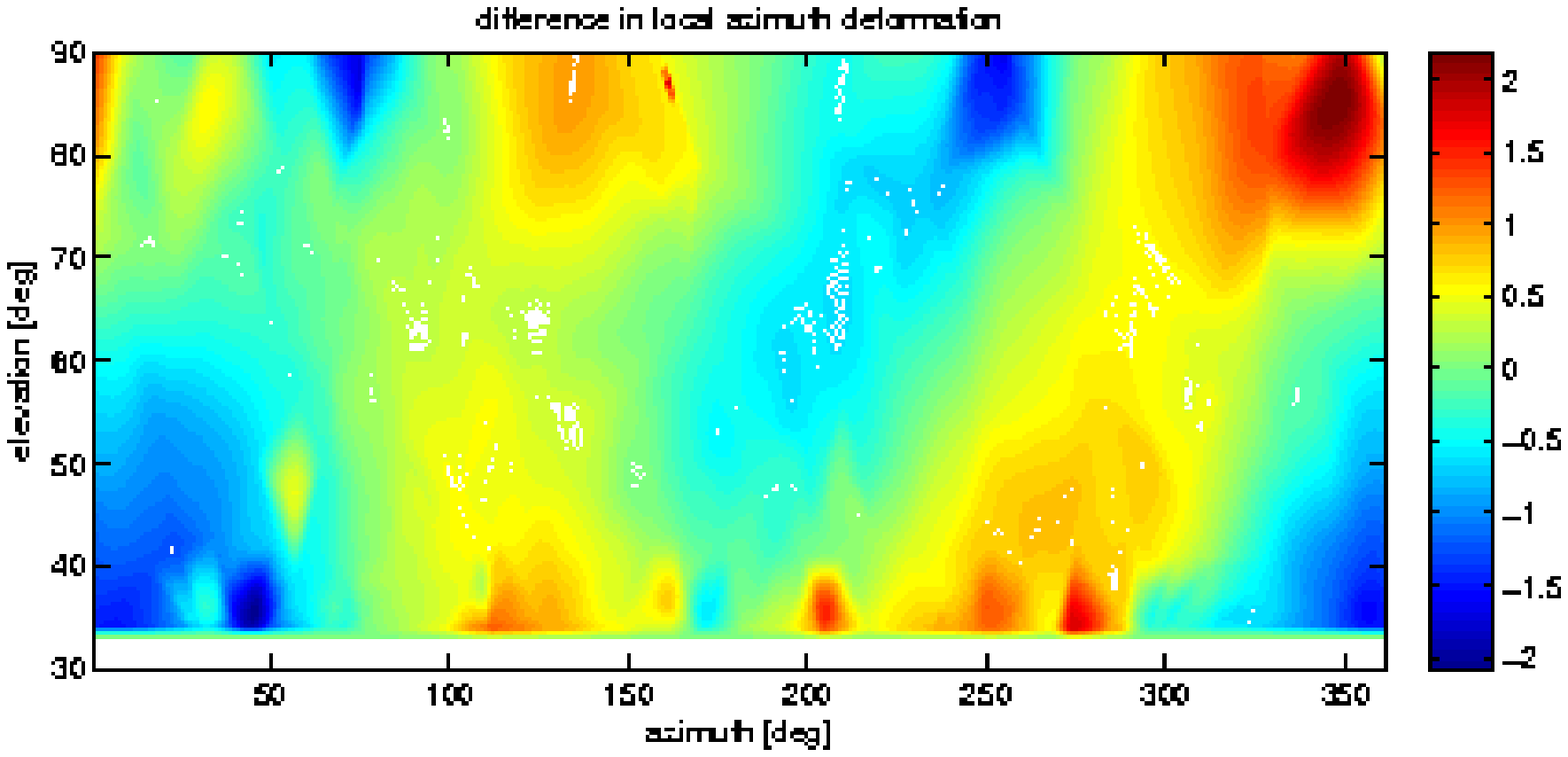}
\includegraphics[scale=0.39]{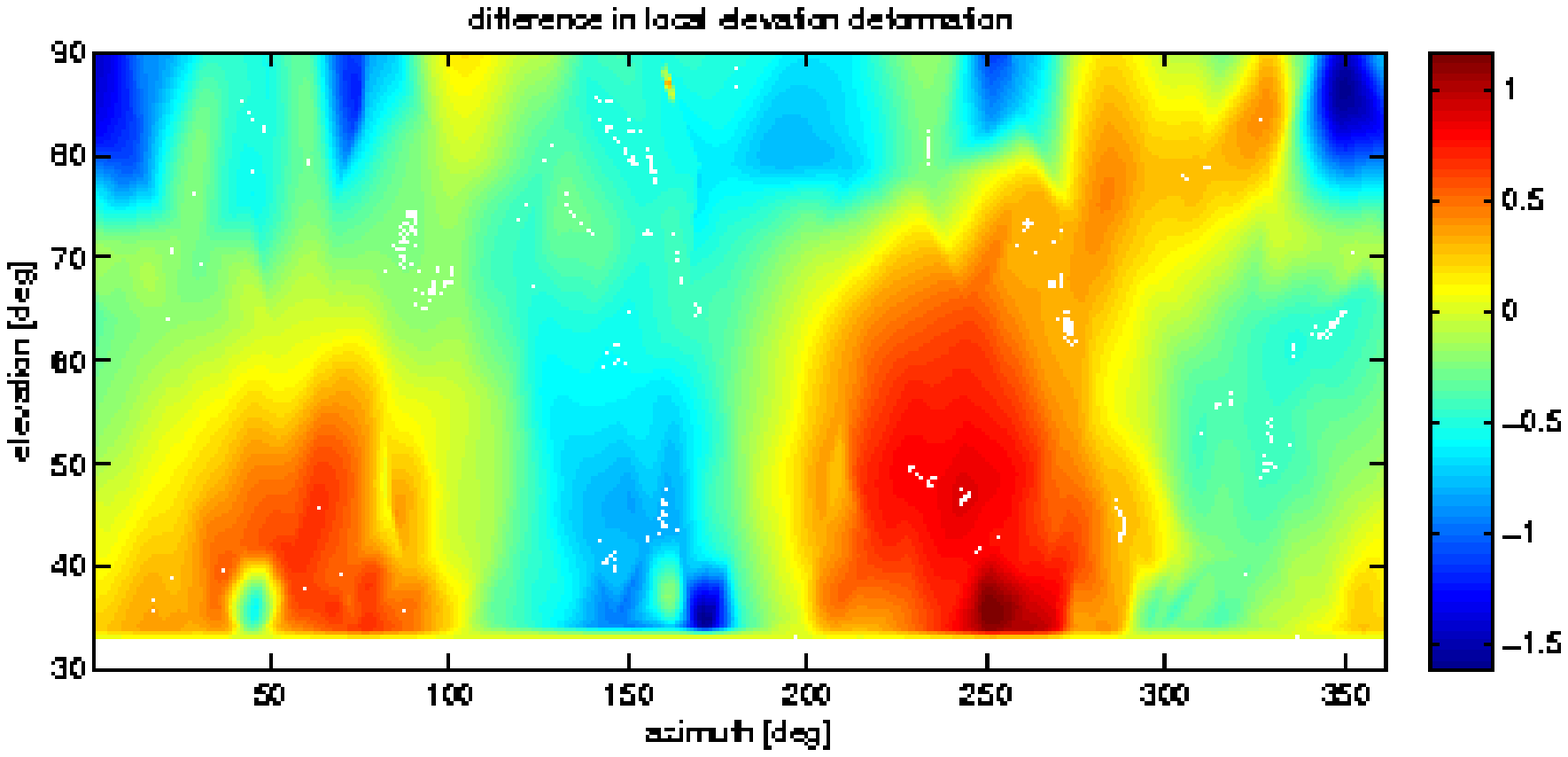}
\caption{\label{check}The local azimuth and elevation deformation errors for OT1 and OT2. Units are in arcmin.
The bottom panels are the difference between the two local deformations, showing a similar structure as
in the Figs.\ref{ot12_az_combined} and \ref{ot12_el_combined}. 
}
\end{center}
\end{figure}
Fig.\ref{separation} shows the result for the 4 fitting parameters. 
Even after removing the platform deformation component, the azimuth and elevation errors show similar 
dominating structures as in the original data sets from OT1 and OT2. The overall positive azimuth error 
is likely due to a remaining cone mis-orientation with respect to north.
The existence of this real dominating structure also explains why the deformation corrected pointing
errors are not random. The newly derived values for  $p_{az}$ and  $p_{el}$ are then 
used for the IT. 
The typical local deformation introduces azimuth and elevation pointing errors between $-1'$ and $+1'$.
The deformation corrected pointing errors are between $-1'$ and up to $2-3'$.\\
As a consistency check we show the separated azimuth and elevation deformation errors for OT1 and OT2
in Fig.\ref{check}. The difference between the local deformation errors between OT1 and OT2 are then 
to be compared with the difference between the raw OT1 and OT2 errors without any modeling (error 
separation) in the bottom panels in the Figs.\ref{ot12_az_combined} and \ref{ot12_el_combined}.
The difference plots indeed agree very well.

\subsection{Pointing limit, phase error and array efficiency}   \label{efficiency}

\begin{center}
\begin{figure}
\includegraphics[scale=0.65]{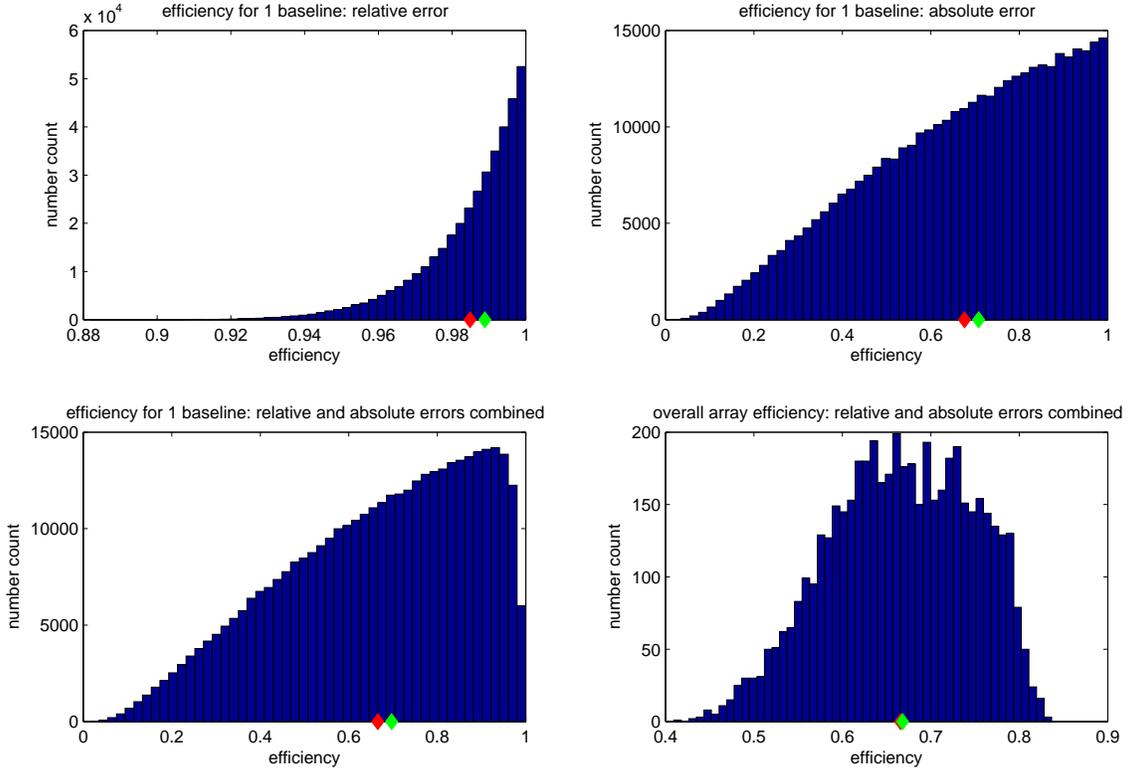}
\caption{\label{13_overall}Efficiencies from a Monte-Carlo simulation for 10.000 realizations 
for an initial set of error intervals: $d\in [-1,1]$, $m\in [-1,1]$, $p\in [-1,1]$.
Top panels: separated relative and absolute errors for one single baseline. Bottom panels: relative
and absolute errors multiplied for a single baseline and for the entire array. 
The average overall efficiency is $\sim 67\%$.
The mean and median values are shown with a red and green marker, respectively.}
\end{figure}
\end{center}

\begin{figure}
\begin{center}
\includegraphics[scale=0.35]{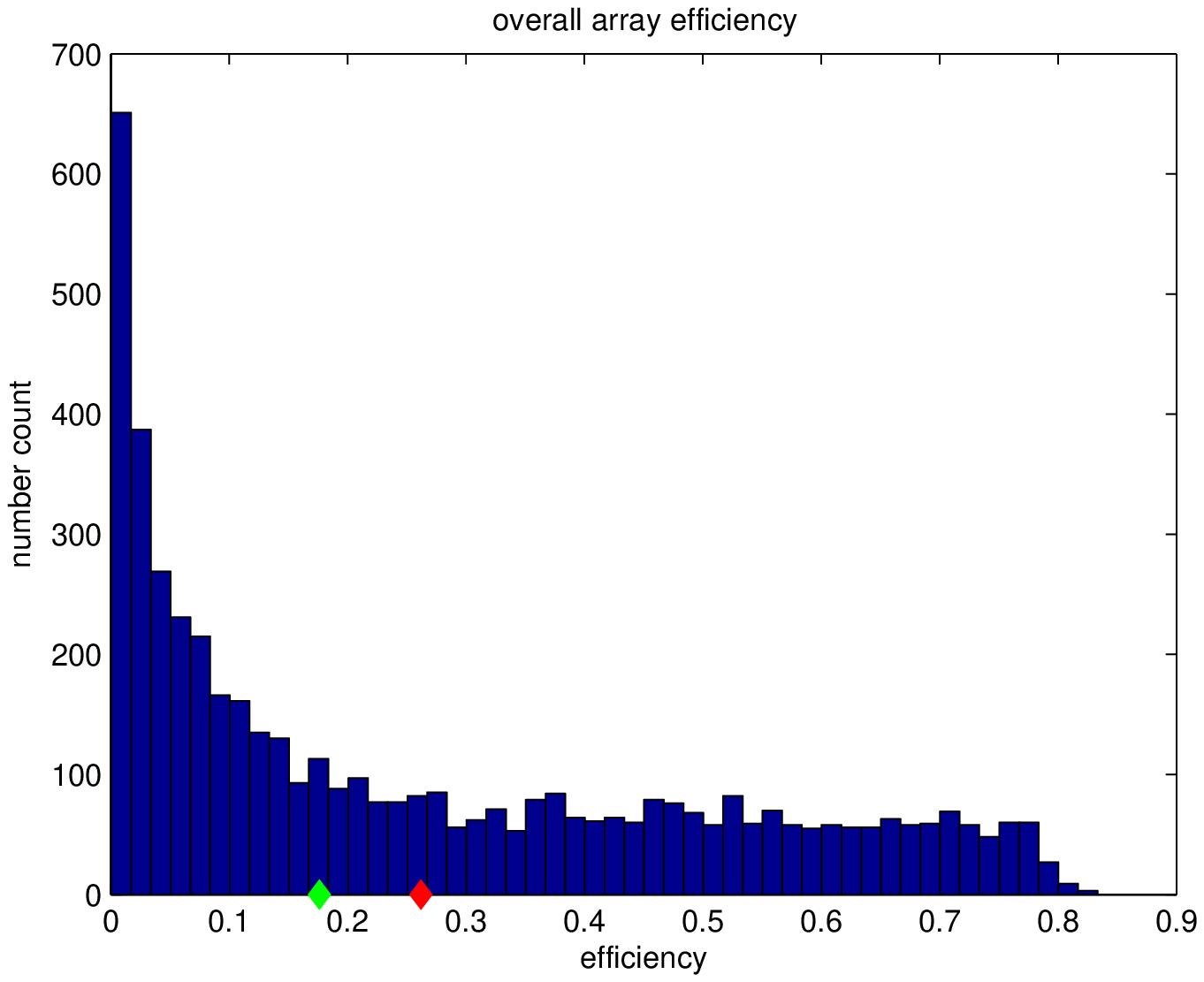}
\includegraphics[scale=0.35]{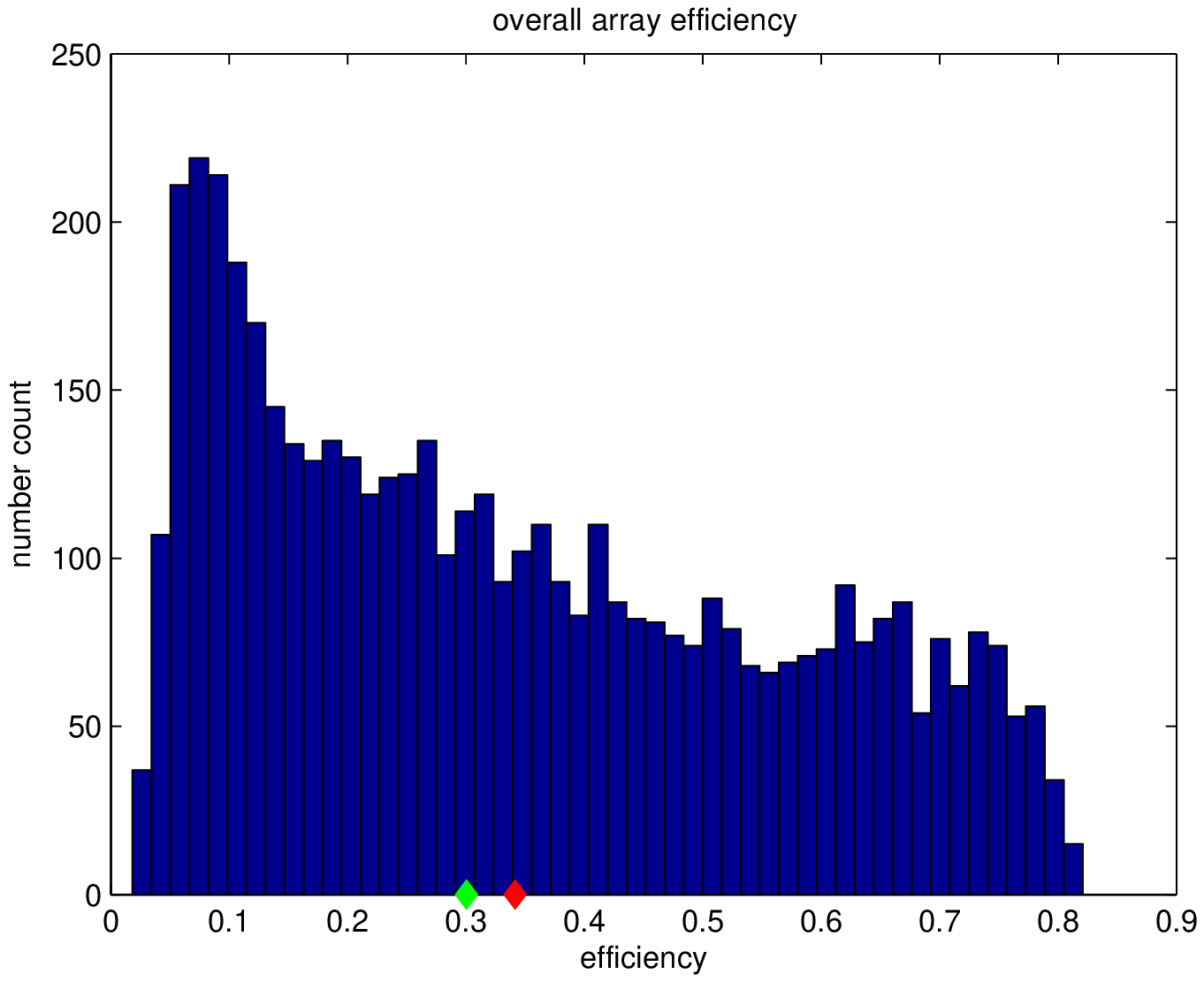}
\includegraphics[scale=0.35]{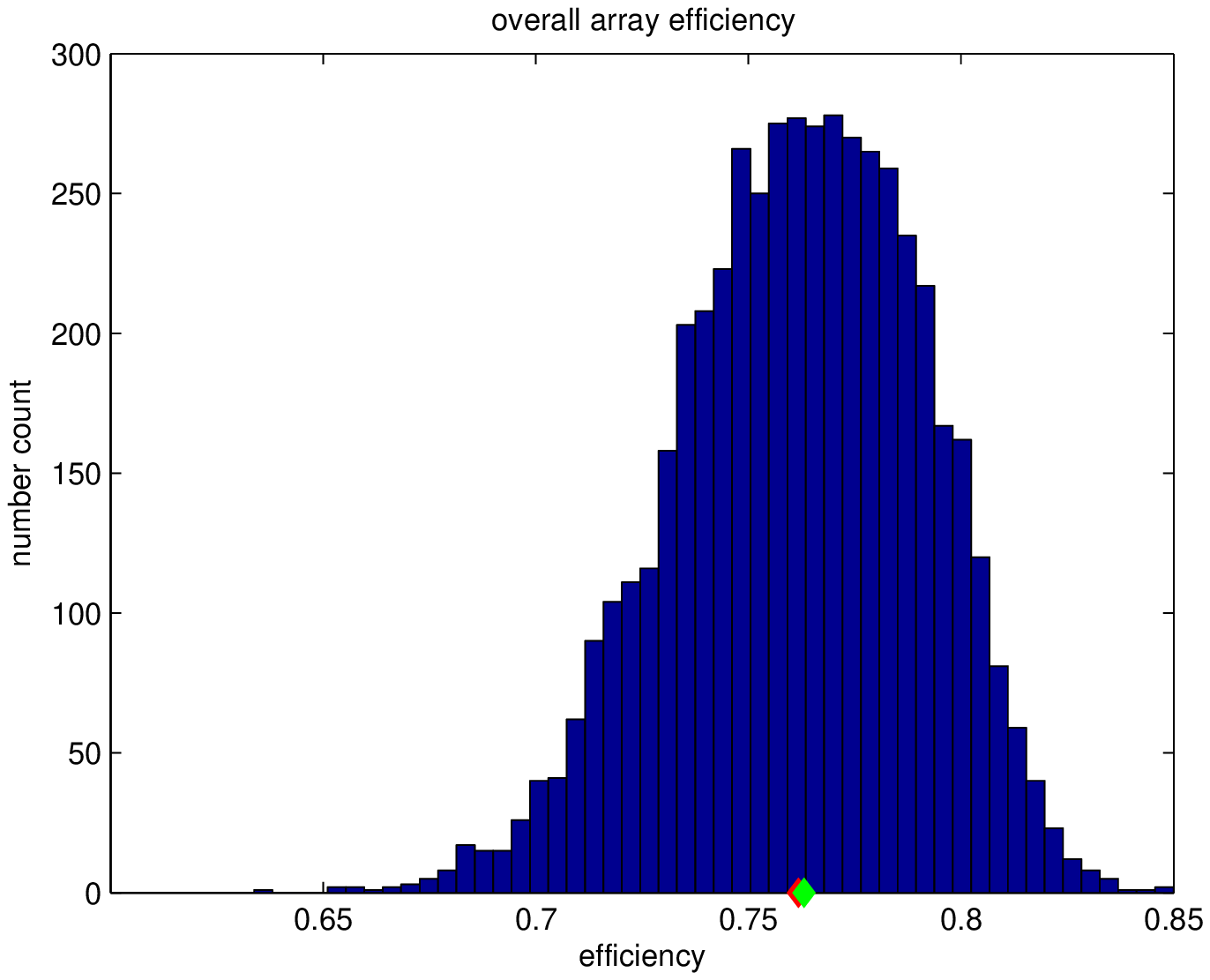}
\includegraphics[scale=0.35]{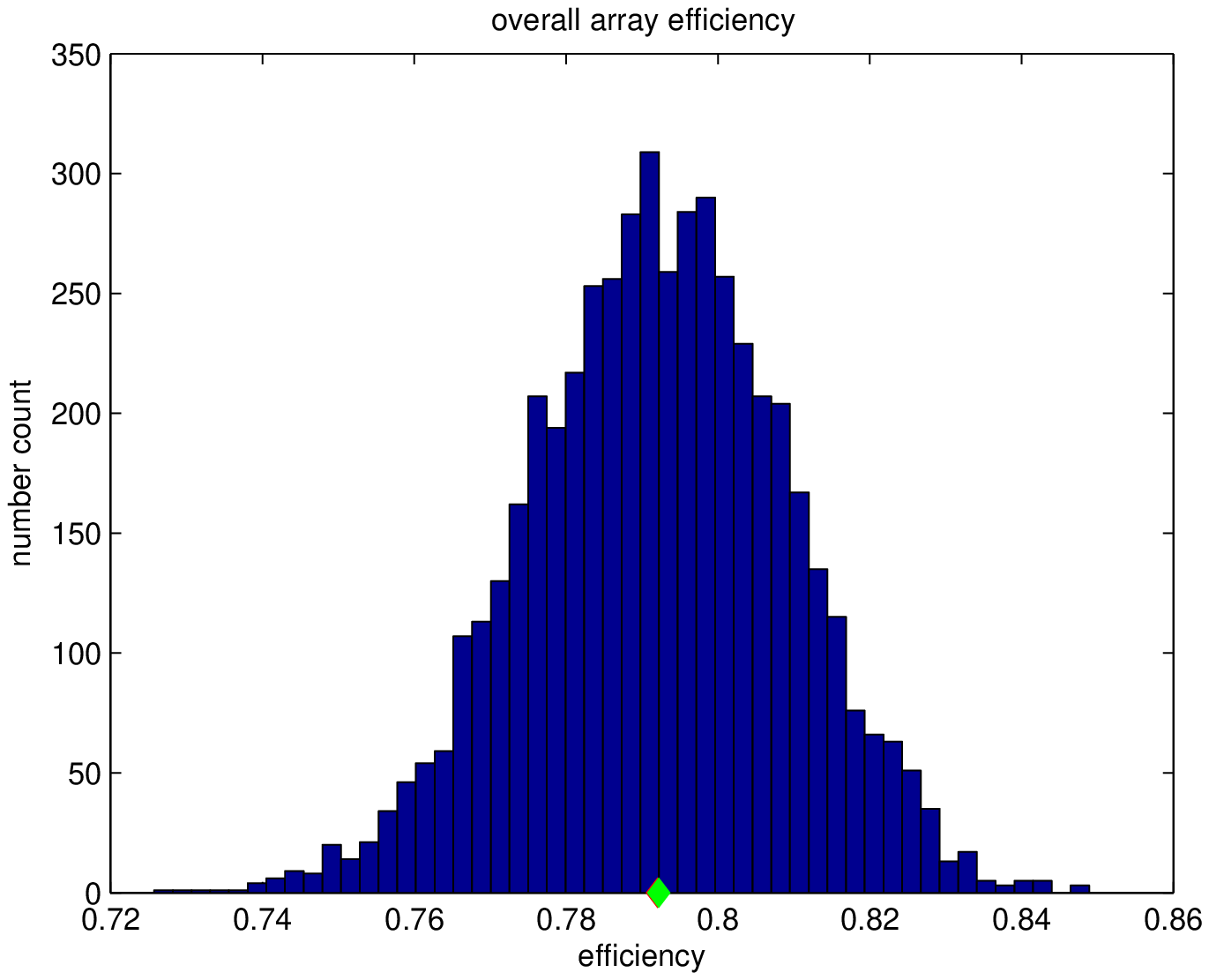}
\includegraphics[scale=0.35]{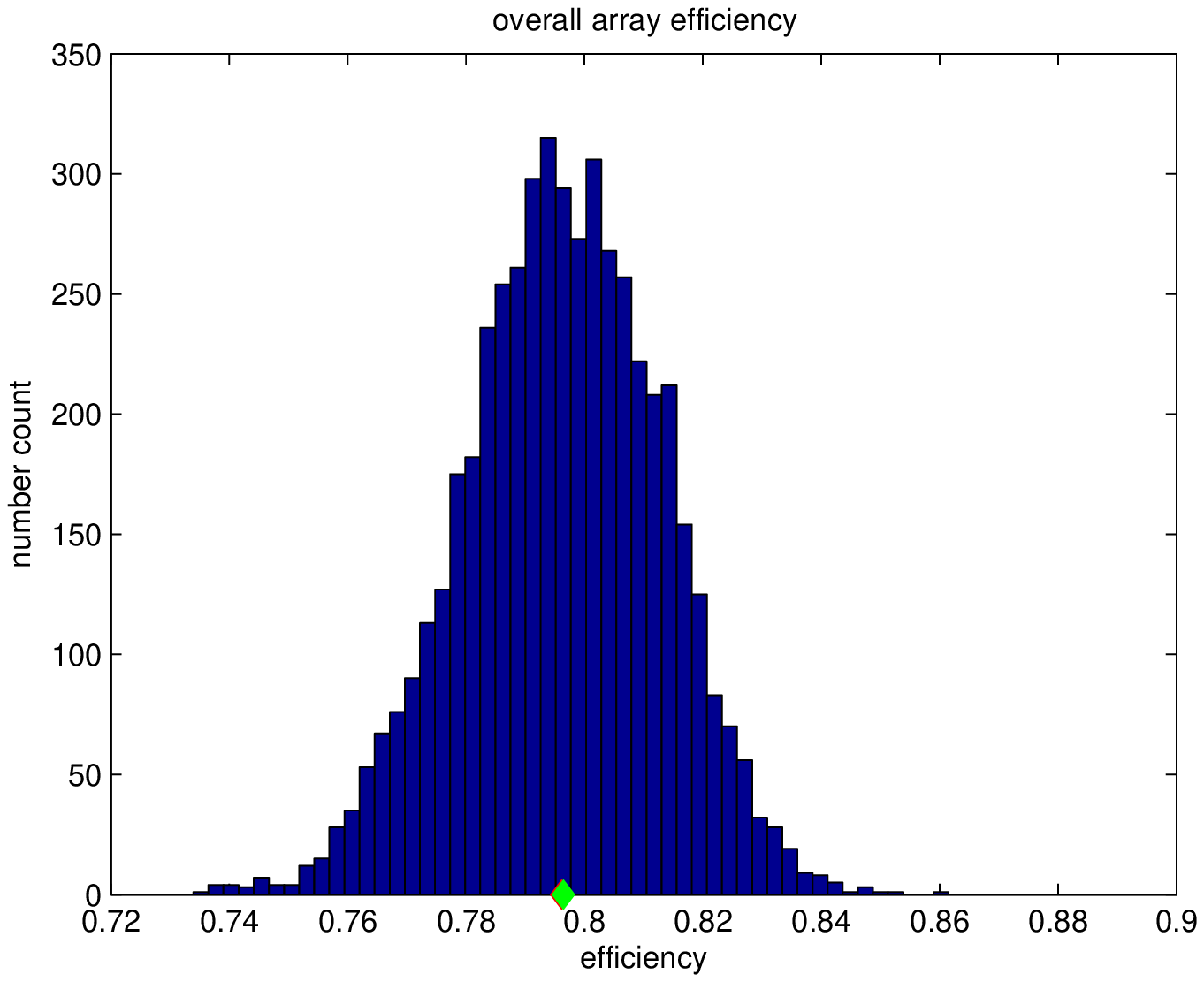}
\includegraphics[scale=0.35]{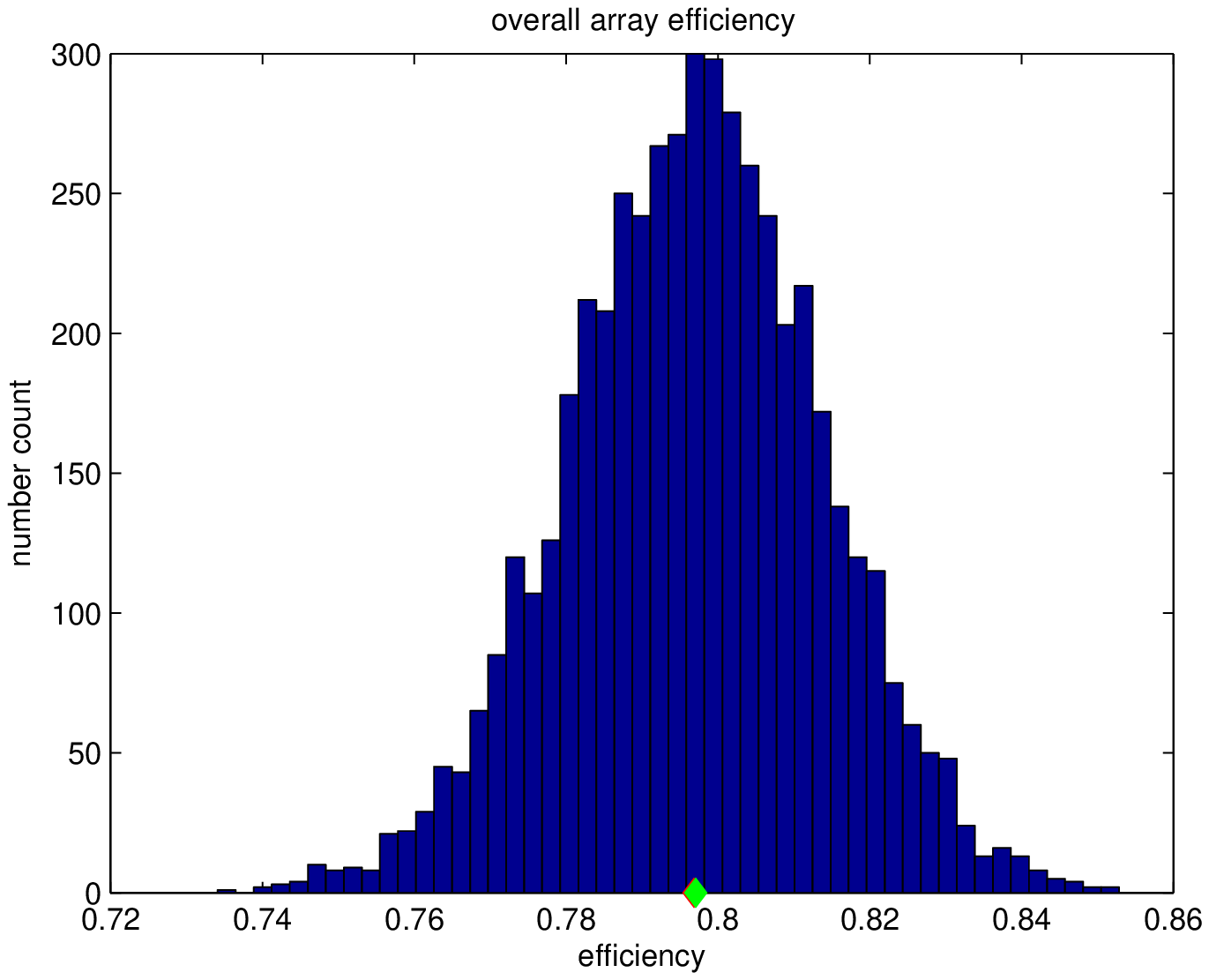}
\caption{\label{13_eff_pointing}Overall efficiencies for successively reduced pointing error intervals,
from left to right, top to bottom:  $d\in [-1,4],[-1,3],[-0.5,0.5], [-0.2,0.2],[-0.1,0.1],[-0.05,0.05]$.
The mean and median values are shown with a red and green marker, respectively. }
\end{center}
\end{figure}

The platform deformation (and any other local error) complicates the definition and the 
extraction of the mount pointing error. The question rises how far the pointing accuracy can 
be improved before the local deformations start to dominate.  
In order to answer this question we quantify the array efficiency loss resulting from the mount pointing
error and the local platform deformation. For completeness we also include a  
radio misalignment error for each baseline (each antenna pair).
For each antenna pair we calculate a relative and an absolute error, $\delta_{rel}$ and $\delta_{abs}$, 
respectively.
$\delta_{rel}$ measures the shift in the overlap of two antenna primary beams
(relative antenna misalignment) due to any local irregularities, whereas $\delta_{abs}$ quantifies the 
(absolute) synthesized beam shift for each antenna pair with respect to a perfect pointing axis (Fig.\ref{errors}, 
right hand side):

\begin{eqnarray}
\delta_{rel}&=&\sqrt{((d_{1,az}-d_{2,az})+(m_{1,az}-m_{2,az}))^2+((d_{1,el}-d_{2,el})+(m_{1,el}-m_{2,el}))^2},\\
\delta_{abs}&=&\sqrt{\left(\frac{d_{1,az}+d_{2,az}}{2}+\frac{m_{1,az}+m_{2,az}}{2}+p_{az}\right)^2+
\left(\frac{d_{1,el}+d_{2,el}}{2}+\frac{m_{1,el}+m_{2,el}}{2}+p_{el}\right)^2}, \label{delta_abs}
\end{eqnarray}
where $d_k$ and $m_k$ are the platform deformation and the misalignment error, respectively, 
for $k=1,2$ for each 
antenna pair, each having an azimuth and elevation error component. $p_{az}$ and $p_{el}$ are the 
mount azimuth and elevation pointing error, respectively.\\
We then further define the relative efficiency, $\eta_{rel}$, per antenna pair (i.e. per correlation) 
as the square root of the product of their
primary Gaussian beams with an overlap error $\delta_{rel}$
(primary synthesized beam). For the absolute array efficiency, $\eta_{abs}$,
additionally an array synthesized beam product with an overlap error $\delta_{abs}$ is multiplied with $\eta_{rel}$
and then averaged over the total number of array baselines $n_{base}$. This means that the efficiency is 
specifically calculated for a source with an extension comparable to the synthesized beam: 

\begin{eqnarray}
\eta_{rel}&=&\frac{1}{N_P}\int_{dV}\sqrt{\mathcal{A}(r,\sigma_P,0)\cdot \mathcal{A}(r,\sigma_P,\delta_{err})}\,d^3r, \\
\eta_{abs}&=&\frac{1}{n_{base}}\sum^{n_{base}}_{i=1}\frac{1}{N_P}\cdot\frac{1}{N_s}
\cdot \int_{dV}\sqrt{\mathcal{A}(r,\sigma_P,0)\cdot \mathcal{A}(r,\sigma_P,\delta_{err_i})}\,d^3r\cdot 
\int_{dV}\mathcal{A}(r,\sigma_s,0)\cdot \mathcal{A}(r,\sigma_s,\delta_{abs_i})\,d^3r,
\end{eqnarray}
where $\mathcal{A}(r,\sigma,\mu)=\frac{1}{\sigma\sqrt{2\pi}}\cdot \exp(-\frac{(r-\mu)^2}{2\sigma^2})$ is a 
Gaussian beam approximation. $\sigma_P$ and $\sigma_s$ are the primary antenna and synthesized beam FWHM,
 respectively. $N_P$ and $N_s$ are their normalization 
factors.\\
In order to estimate the efficiency we run a Monte-Carlo simulation where we assume all the errors 
($d$, $m$, $p$) to be uniformly distributed in their corresponding measured error intervals\footnote
{Strictly speaking this assumption is not entirely correct because some of the errors are not 
independent, but correlated through the regular deformation patterns.
}.
Fig.\ref{13_overall} illustrates the different efficiencies for an initial set of error intervals (in arcmin):
$d\in [-1,1]$, $m\in [-1,1]$, $p\in [-1,1]$, each having an azimuth and an elevation component. For the 
following discussion we consider the error intervals for $d$ and $m$ to be fixed, $d$ being a hard limit
and $m$ being considered an achievable limit after the radio alignment tuning. All the three errors
can add up or compensate each other, where  $\delta_{abs}$ is more important when trying to improve
the efficiency. $\delta_{rel}\sim 0.98$ is not of a concern at all. In a series of Monte-Carlo simulations
$\eta_{abs}$ has been calculated for successively smaller error intervals for $p$, 
shown in Fig.\ref{13_eff_pointing}.

From the Figs.\ref{13_overall} and \ref{13_eff_pointing} it is clear that the overall efficiency 
reaches a first limit, average and median $\sim 0.67$, when all the error components have the 
same error interval, which is set by $d\in[-1,1]$. However, successively smaller pointing errors 
can still improve the efficiency by about up to $10\%$ due to the random summation and compensation
of the errors. The final limit is reached at $\sim 0.8$ where smaller errors are absorbed and dominated by 
the largest error interval. A similar improvement ($\sim$10\%) would be possible if $m$ or $d$ 
could be made smaller. An additional overall factor of about 0.9 will slightly reduce all these numbers, assuming
that the distorted synthesized beam can be restored to about 90\% with the phase correction model.

\section{Summary and Conclusion}

We have demonstrated that the local platform deformation error can be isolated from the pointing 
error with the help of two OTs. The extracted deformation error is in the range $-1'$ to $+1'$ which is 
consistent with the photogrammetry data. The verification of an improved pointing accuracy is, however,
non-trivial, because the platform deformation component will always be present on the ccd images.
Therefore, one always has to rely on the modeling. 
Aiming for a pointing accuracy of $\sim 0.5'$ rms ($\eta_{abs}\sim 0.76$) seems reasonable. Even
smaller pointing errors improve the efficiency only marginally. 
Similarly, the platform deformation induced phase error is corrected with the photogrammetry data, 
and the synthesized beam amplitude is restored to 90\% or more. \\
Finally, one has to ask whether the separation of platform and pointing error is needed at all.
Assuming that the errors are not separated and the derived pointing errors are then reduced to zero, 
one still ends up with an effective pointing error ($p_{az}$ and $p_{el}$ in Eq.(\ref{delta_abs}))
in the range $[-1,1]$ due to the over-/under-compensated platform deformation error. $\delta_{abs}$
can therefore not be further reduced and the efficiency is limited to 67\%. 
For large pointing errors ($>1'$) the separation becomes unnecessary. Going beyond $\eta_{abs}\sim 0.67$
requires a better effective pointing and a separation of the error components. In any case, a  phase correction
model is needed to restore the synthesized beam.

\acknowledgments 
Capital and operational funding for AMiBA came from the Ministry of Education and the 
National Science Council as 
part of the 
Cosmology and Particle Astrophysics
(CosPA) initiative. Matching operational funding also came in the form of an Academia Sinica Key Project.

\section*{REFERENCES}

\bibliography{report}   

\bibliographystyle{spiebib}   

\indent\hspace{0.42cm} 1. M.Birkinshaw, 1999. Phys. Rep. 310, 97

2. J.A. Peacock, 1999. in: {\it Cosmological Physics}, Cambridge University Press

3. P. Koch, Ph. Raffin, J.-H.P. Wu, M.-T. Chen, T.-H. Chieuh, P.T.P. Ho, C.-W. Huang, Y.-D. Huang Y.-W. Liao, 
K.-Y. Lin, 
G.-C. Liu, H. Nishioka, C.-L. Ong, K. Umetsu, F.-C. Wang, S.-K. Wong, Ch. Granet,
 "0.6m antennae for the AMiBA interferometry array",
in: Proceedings of The European Conference on Antennas and Propagation: EuCAP 2006 (ESA SP-626)
Editors: H. Lacoste \& L. Ouwehand. Published on CDROM., p.668.1, 2006

4. P. Raffin, P. Koch, Y.-D. Huang, C.-H. Chang, J. Chang, M.-T. Chen, K.-Y. Chen, 
P.T.P. Ho, C.-W. Huang, F. Ibanez-Roman, 
H.-M. Jiang, M. Kesteven, K.-Y. Lin, G.-C. Liu, H. Nishioka, K. Umetsu,
 "Progress of the Array for Microwave Background Anisotropy (AMiBA)",
in: Optomechanical Technologies for Astronomy,
Proc.of SPIE 6273, pp 468-481, 2006

5. Y.-D. Huang, P. Raffin, M.-T. Chen, P. Altamirano, P. Oshiro,
"Photogrammetry measurement of the AMiBA 6-meter platform",
in: Ground-based and Airborne Telescopes, Proc.of SPIE, 2008 (upcoming)

6. P.T.P. Ho, M.-T. Chen, T.-D. Chiueh, T.-H. Chiueh, T.-H. Chu, 
H.-M. Jiang, P. Koch, D. Kubo, C.-T. Li, M. Kesteven, K.-Y. Lo, 
C.-J. Ma, R.N. Martin, K.-W. Ng, H. Nishioka, F. Patt, J.B. Peterson, P. Raffin, H. Wang, Y.-J. Hwang, K. Umetsu, 
and J.-H.P.Wu, "The AMiBA Project", Modern Physics Letters A, 19 (2004) p.993

 7. P. Raffin, R.N. Martin, Y.-D. Huang, F. Patt, R.C. Romeo, M.-T. Chen, and J.S. Kingsley, "CFRP Platform and Hexapod 
Mount for the Array of MIcrowave Background Anisotropy (AMiBA)", SPIE Europe International Symposium, Proc. 
Astronomical Structures and Mechanisms Technology,
Glasgow, Scotland, UK, 2004

8. C.-T. Li, C.-C. Han, M.-T. Chen, K. Umetsu, K.-Y. Lin, Y.-D. Huang, W. Wilson, H. Jiang, 
Y.-J. Hwang, S.-W. Chang, C.-H. Chang, J. Chang, P. Martin-Cocher, G.-C. Liu, 
H. Nishioka, P.T.P. Ho, "Initial Operation of the Array of 
Microwave Background Anisotropy (AMiBA)",
 in: Millimeter and Submillimeter Detectors and Instrumentation for Astronomy,
Proc.of SPIE 6275, pp 487-498, 2006

9. K.-Y. Lin, C.-T. Li, J.-H.P. Wu, K. Umetsu, P.M. Koch, G.-C. Liu,
H. Nishioka, C.-W. Huang, Y.-W. Liao, F.-C. Wang, P. Altamirano,
D. Kubo, C.-C. Han, S.-W. Chang, C.-H. Chang, Y.-D. Huang,
P. Raffin, P. Oshiro, H.-M. Jian, M.-T. Chen. and P.T.P. Ho,
"AMiBA First-year Observation",
in: Ground-based and Airborne Telescopes, Proc.of SPIE, 2008 (upcoming)   

\end{document}